\documentclass[twocolumn]{aastex631}
\usepackage{amsmath}
\usepackage{mathtools}
\usepackage{color}
\usepackage{hyperref}

\newcommand{\specnotation}[2]{\ensuremath{\rm #1 \, {\scriptstyle #2}}}
\newcommand{\OVI}{\specnotation{O}{VI}}
\newcommand{\atomH}{\specnotation{H}{I}}

\newcommand{\Mstar}{{\ensuremath{M_*}}}
\newcommand{\CGMsq}{CGM$^2$}
\newcommand{\RCCc}{\ensuremath{\rm R_{\rm 200c}}}
\newcommand{\MCCc}{\ensuremath{\rm M_{\rm 200c}}}
\newcommand{\TCCc}{\ensuremath{\rm T_{\rm 200c}}}
\newcommand{\Msun}{\ensuremath{M_{\odot}}}
\newcommand{\dd}{\,{\rm d}}
\newcommand{\logten}{{\rm log}_{10}}

\newcommand{\NOVI}{\ensuremath{N_{\OVI}}}
\newcommand{\nOVI}{\ensuremath{n_{\OVI}}}
\newcommand{\ApNOVI}[1]{\ensuremath{\langle \NOVI \rangle_{#1}}}

\newcommand{\dNdz}{\ensuremath{\dd N / \dd z}}
\newcommand{\cmmt}{cm\ensuremath{^{-2}}}
\newcommand{\IP}{\ensuremath{R_{\perp}}}
\newcommand{\IPmax}{\ensuremath{R_{\perp,max}}}
\newcommand{\lten}{\ensuremath{\rm log_{10}}}
\newcommand{\Nzero}{\mathcal{N}_0}
\newcommand{\Nbase}{\mathcal{N}_b}
\newcommand{\scalelen}{L_s}
\newcommand{\rng}{\text{--}}

\graphicspath{{./}{figures/}}
\defcitealias{Wilde:2021vr}{W21}

\begin{document}

\title{The \CGMsq\ Survey: Circumgalactic \OVI\ From Dwarf to Massive Star-Forming Galaxies}

\author[0000-0003-0789-9939]{Kirill Tchernyshyov}
\affiliation{Department of Astronomy, University of Washington, Seattle, WA 98195, USA}
\author[0000-0002-0355-0134]{Jessica K. Werk}
\affiliation{Department of Astronomy, University of Washington, Seattle, WA 98195, USA}
\author[0000-0003-1980-364X]{Matthew C. Wilde}
\affiliation{Department of Astronomy, University of Washington, Seattle, WA 98195, USA}
\author[0000-0002-7738-6875]{J. Xavier Prochaska}
\affiliation{University of California, Santa Cruz; 1156 High Street, Santa Cruz, CA 95064, USA}
\affiliation{Kavli Institute for the Physics and Mathematics of the Universe (Kavli IPMU) The University of Tokyo; 5-1-5 Kashiwanoha, Kashiwa, 277-8583, Japan}
\author[0000-0002-1218-640X]{Todd M. Tripp}
\affiliation{Department of Astronomy, University of Massachusetts, 710 North Pleasant Street, Amherst, MA 01003-9305, USA}
\author[0000-0002-1979-2197]{Joseph N. Burchett}
\affiliation{University of California, Santa Cruz; 1156 High Street, Santa Cruz, CA 95064, USA}
\affiliation{Department of Astronomy, New Mexico State University, PO Box 30001, MSC 4500, Las Cruces, NM 88001, USA}

\author[0000-0002-3120-7173]{Rongmon Bordoloi}
\affiliation{Department of Physics, North Carolina State University, Raleigh, NC 27695-8202, USA}

\author[0000-0002-2591-3792]{J. Christopher Howk}
\affiliation{Department of Physics, University of Notre Dame, Notre Dame, IN 46556}

\author[0000-0001-9158-0829]{Nicolas Lehner}
\affiliation{Department of Physics, University of Notre Dame, Notre Dame, IN 46556}

\author[0000-0002-7893-1054]{John M. O'Meara}
\affil{W. M. Keck Observatory, 65-1120 Mamalahoa Hwy., Kamuela, HI 96743, USA}

\author[0000-0002-1883-4252]{Nicolas Tejos}
\affil{Instituto de F\'isica, Pontificia Universidad Cat\'olica de Valpara\'iso, Casilla 4059, Valpara\'iso, Chile}

\author[0000-0002-7982-412X]{Jason Tumlinson}
\affil{Space Telescope Science Institute, Baltimore, MD, USA}

\correspondingauthor{Kirill Tchernyshyov}
\email{ktcherny@gmail.com}

\begin{abstract}
We combine 126 new galaxy-\OVI\ absorber pairs from the \CGMsq\ survey with 123 pairs drawn from the literature to examine the simultaneous dependence of the column density of \OVI\ absorbers (\NOVI) on galaxy stellar mass, star formation rate, and impact parameter.
The combined sample consists of 249 galaxy-\OVI\ absorber pairs covering $z=0\rng0.6$, with host galaxy stellar masses $\Mstar=10^{7.8}\rng10^{11.2}$ \Msun\ and galaxy-absorber impact parameters  $\IP=0\rng400$ proper kiloparsecs.
In this work, we focus on the variation of \NOVI\ with galaxy mass and impact parameter among the star-forming galaxies in the sample.
We find that the average \NOVI\ within one virial radius of a star-forming galaxy is greatest for star-forming galaxies with $\Mstar=10^{9.2} \rng 10^{10}$ \Msun.
Star-forming galaxies with \Mstar\ between $10^{8}$ and $10^{11.2}$ \Msun\ can explain most \OVI\ systems with column densities greater than 10$^{13.5}$ \cmmt.
60\% of the \OVI\ mass associated with a star-forming galaxy is found within one virial radius and 35\% is found between one and two virial radii.
In general, we find that some departure from hydrostatic equilibrium in the CGM is necessary to reproduce the observed \OVI\ amount, galaxy mass dependence, and extent.
Our measurements serve as a test set for CGM models over a broad range of host galaxy masses.
\end{abstract}

\section{Introduction}
\label{sec:intro}
The diffuse gaseous halos around galaxies, their circumgalactic media or CGM, are as important in galaxy evolution as the stars and interstellar gas that are inside galaxies.
The CGM is the site from which galaxies draw gas to fuel continued star formation and into which feedback ejects gas and energy.
It may even play a more active role in galaxy evolution: the CGM's thermodynamic and hydrodynamic conditions may regulate the accretion of intergalactic gas into the galactic system, the supply of gas to the host galaxy from the CGM, and the pressure balance in the host galaxy's ISM \citep{Voit:2019tv,Terrazas:2020wg,Oppenheimer:2020tx,Davies:2020wd,Zinger:2020vm}.

The goal of this work is to better understand how the column density of the ion \OVI\ (O$^{+5}$) in the CGM of a galaxy is connected to the galaxy's properties (e.g., mass and star formation rate).
The purpose of understanding this connection is to help determine what kind of conditions and generating mechanisms are traced by the presence of \OVI.
Multiple plausible mechanisms that correspond to very different conditions can produce \OVI.
Comparing the CGM conditions inferred from \OVI\ to the properties of the host galaxies over a wide range in galaxy mass can provide insight on the origin of \OVI\ and the role of the CGM in galaxy evolution.
At the most basic level, in CGM regions where the ionizing radiation field is dominated by the unattenuated cosmic ultraviolet background  \citep{Haardt:2012ti,Faucher-Giguere:2020wt}, the ionization balance of \OVI-traced gas can be set by either collisional ionization or photoionization.
In equilibrated gas, collisional ionization gives a high \OVI\ fraction in gas with number density $n_{\rm H}>10^{-4}$ cm$^{-3}$ and temperature $T\approx 10^{5.5}$ K while photoionization gives a high \OVI\ fraction in gas with $n_{\rm H}\sim 10^{-6}\rng10^{-4}$ cm$^{-3}$ and $T\sim 10^{4}$ K \citep[e.g.,][]{Stern:2018to}.
However, \OVI\ can also be detected in gas that is out of ionization equilibrium \citep{Gnat:2007tt,Oppenheimer:2013vs}, in which case the instantaneous physical conditions may not reflect the full thermodynamic history of the gas.
At a higher level, there are many possible mechanisms for generating one or the other of these sets of \OVI-rich conditions.
A non-exhaustive list of possible mechanisms includes shock ionization \citep{Dopita:1996wz}, turbulent mixing layers \citep{Slavin:1993ve,Tan:2021tc}, conductive interfaces \citep{Borkowski:1990ut,Gnat:2010wu}, static radiative non-equilibrium cooling \citep{Edgar:1986tr,Gnat:2007tt,Oppenheimer:2013vs}, radiative cooling flows \citep{Heckman:2002wz,Wakker:2012vm,Bordoloi:2017vr,McQuinn:2018vk,Stern:2019tp}, and supernova-driven outflows \citep{Thompson:2016vp,Li:2020wn}.
These mechanisms can be categorized as involving (1) ambient halo gas, (2) outflows from the host galaxy, or (3) inflows towards and onto the host galaxy.

All three categories of mechanisms are no doubt in action to varying extents.
The question is their relative contribution as a function of the properties of a CGM host galaxy.
This question has been thoroughly explored in the theoretical literature, producing hypotheses centering each of the three categories of mechanism.
A meta-conclusion we draw from this work is that it is possible to create many plausible and effective models for \OVI\ in the CGM of galaxies \emph{in a narrow mass range}.
For example, \citet{Faerman:2020wc} (collisionally ionized ambient gas), \citet{Li:2020wn} (collisionally ionized outflows), and \citet{Stern:2018to} (photoionized inflows) are based on qualitatively different assumptions for how most of the \OVI\ is generated, but agree with the set of \OVI\ measurements that they all reference \citep{Werk:2013uj,Johnson:2015tj}; both the models and the measurements are of roughly Milky Way-mass galaxies.
In order to learn what mechanisms are actually responsible for \OVI\ (and hence what \OVI\ measurements mean), it is necessary to consistently make predictions for \OVI\ in the CGM of galaxies over a wide range of masses, star formation rates (SFRs), and other properties and to confront these predictions with measurements.

This necessary confrontation has been prevented by a shortage of observations.
The available measurements of \OVI\ around galaxies in the literature are too sparse to create statistically useful sub-samples over narrow ranges simultaneously in three key properties: galaxy stellar mass (\Mstar), impact parameter (\IP), and star formation rate.
In this work, we combine measurements from the literature with new galaxy-\OVI\ absorber pairs drawn from the \CGMsq\ survey (\citealt{Wilde:2021vr}, referred to below as \citetalias{Wilde:2021vr}) to create a sample large enough to allow simultaneous slicing in all three of these properties.
\CGMsq\ is a deep galaxy redshift survey of 22 fields with far ultraviolet quasar spectroscopy from the COS-Halos and COS-Dwarfs \citep{Tumlinson:2013vs,Bordoloi:2014uz} surveys.
\CGMsq\ contains 126 unique, previously unpublished galaxy-\OVI\ absorber pairs with \IP$<400$ kpc at $z=0.1\rng0.6$ and includes galaxies with stellar mass $\Mstar\sim10^{8}\rng10^{11}$ \Msun, i.e., sub-Small Magellanic Cloud dwarfs to super-$L^{*}$ galaxies.

In \S \ref{sec:data}, we combine galaxy and \OVI\ absorption data from \CGMsq\ and from literature sources to create a galaxy-absorber pair sample.
In \S \ref{sec:analysis}, we investigate how the column density of \OVI\ around star-forming galaxies depends on galaxy mass and the galaxy-absorber impact parameter.
We discuss some implications of our findings in \S \ref{sec:discussion} and summarize our results in \S \ref{sec:conclusion}.
We assume a flat-universe $\Lambda$CDM cosmology with $H_0=67.8$ km s$^{-1}$ Mpc$^{-1}$ and $\Omega_m=0.308$ \citep{Planck-Collaboration:2016tp}.
Stellar masses are derived assuming a \citet{Chabrier:2003ta} initial mass function.

\section{Data}
\label{sec:data}

Our study requires both galaxy and QSO spectroscopy to uniquely associate specific galaxies with \OVI{~$\lambda\lambda$ 1031,1037\AA} absorption at precisely-measured redshifts.
Other galaxy properties, such as stellar masses and SFRs, are derived by fitting multi-band photometric measurements using the spectral energy distribution fitting package \texttt{CIGALE}\footnote{Version 2020.0, \citet{Boquien:2019wx}}.
\OVI\ absorbers are measured from medium-resolution far ultraviolet spectra of quasars.
We combine our own measurements with ones drawn from earlier surveys of galaxies and \OVI\ absorbers.
The surveys we use are listed in Table \ref{tab:galaxy-samples}.
Our procedure for analyzing the \CGMsq\ dataset and incorporating measurements from the literature are described in the remainder of this section.

\subsection{Galaxy surveys}
\label{sec:data:galaxies}
The first part of the dataset is a collection of properties---redshifts, masses, and a classification as star-forming or quiescent---for galaxies in QSO fields.

\begin{deluxetable*}{lcccc}
\tabletypesize{\footnotesize}
\tablecolumns{5}
\tablewidth{0pt}
\tablecaption{ Galaxy-absorber surveys included in our sample \label{tab:galaxy-samples}}
\tablehead{\colhead{Name} & \colhead{Number} & \colhead{Galaxy Treatment} & \colhead{\NOVI\ treatment} & References
}
\startdata
COS-Halos & 37 & D & A & \citet{Werk:2012ug,Werk:2013uj}\\
eCGM & 33 & D & A & \citet{Johnson:2015tj}\\
Johnson+2017 & 11 & A & D & \citet{Johnson:2017tp}\\
Keeney+2017 & 13 & A & A & \citet{Keeney:2017wk}\\
Keeney+2018 & 18 & D & A/D & \citet{Keeney:2018wp,Danforth:2016uj,Stocke:2019vr}\\
COS-LRG & 7 & D & A & \citet{Chen:2018tu,Zahedy:2019vq}\\
RDR & 2 & D & A & \citet{Berg:2019wq}\\
QSAGE & 2 & A & A & \citet{Bielby:2019vv}\\
\CGMsq\ & 126 & A & D & \citet{Wilde:2021vr}\\
\enddata
\tablecomments{Galaxy redshift and \OVI\ absorption surveys used in this work. Number---The number of unique galaxy-absorber pairs included in our sample. Galaxy Treatment---\emph{A} means we (A)dopt the stellar masses and photometric star formation rates given by the source, \emph{D} means we (D)erive these quantities ourselves using public photometric surveys and \texttt{CIGALE}. \NOVI\ treatment---\emph{A} means we (A)dopt column densities and upper limits from the literature source, \emph{D} means we (D)erive these quantities ourselves from the quasar spectra. Keeney+2018 has both A and D in the \NOVI\ treatment column because some of their galaxies without detected absorbers do not have published \NOVI\ upper limits; we derive these upper limits ourselves but otherwise adopt values from the references.}
\end{deluxetable*}

\subsubsection{\CGMsq}
\label{sec:data:galaxies:cgmsq}
Our new galaxy data set is drawn from the \CGMsq\ survey (\citetalias{Wilde:2021vr}).
The \CGMsq\ survey provides secure spectroscopic redshifts for 971 galaxies in the foreground of 22 QSOs with {\emph{HST}}/COS spectra.
\citetalias{Wilde:2021vr} describe data acquisition and the analysis procedures in depth; we give a short summary here.
The galaxy spectra were taken using the Gemini-GMOS spectrographs on the Gemini North and Gemini South telescopes \citep{Hook:2004wu,Gimeno:2016ul}.
Galaxy redshifts were inferred from the spectra using a combination of the \texttt{REDROCK}\footnote{\url{https://github.com/desihub/redrock}} redshift fitting code and manually checked using the \texttt{VETRR}\footnote{https://github.com/mattcwilde/vetrr} code.

\citetalias{Wilde:2021vr} used \texttt{CIGALE} to derive galaxy stellar masses and star formation rates from an assemblage of photometry.
The photometry includes measurements in the Gemini $g$ and $i$ pre-mask imaging data, SDSS DR14 $ugriz$, DESI Legacy Imaging Surveys DR8 $grz$, PanSTARRS DR2 $grizy$, and WISE 3.4, 4.6, 12, and 22 $\mu$m bands \citep{Abolfathi:2018vl,Dey:2019uk,Chambers:2016vk,Cutri:2013tn}.
In cases where multiple measurements in similar bands are available, we use all of them.
\citetalias{Wilde:2021vr} used \citet{Bruzual:2003wd} stellar population models built assuming a \citet{Chabrier:2003ta} initial mass function as their \texttt{CIGALE} basis functions.
Other \texttt{CIGALE} galaxy template options are listed in \citetalias{Wilde:2021vr}.
We adopt the \citetalias{Wilde:2021vr} stellar masses and photometric SFRs.

We estimate halo masses using the stellar mass-halo mass relation defined in Table J1 of \citet{Behroozi:2019up}, who use the \citet{Bryan:1998tw} convention for halo masses.
Using the procedure defined in \citet{Hu:2003tf}, we convert their halo masses to the convention in which the average mass density within the halo radius is 200 times the critical density of the universe.
We denote these halo masses and the corresponding virial radii as \MCCc\ and \RCCc.
We use the critical density convention to be consistent with theoretical studies with which we compare our measurements.

We classify galaxies as being star-forming (SF) or quiescent (E) using their specific star formation rates (sSFRs) as determined from photometry by \texttt{CIGALE}.
Galaxies with sSFR$>10^{-10}$ yr$^{-1}$ are classified as SF; galaxies with lower sSFRs are classified as E.
The SFR and sSFR estimates we make using \texttt{CIGALE} are based on a limited range in wavelength and so are quite plausibly inaccurate.
We therefore treat them not as realistic estimates of an actual SFR but as proxies that we use for an empirically motivated classification.

While we do have spectroscopic classifications for the \CGMsq\ galaxies based on the presence of absorption and emission lines in the galaxy spectrum, we do not have access to galaxy spectra for the entire literature sample.
We instead use the \CGMsq\ spectroscopic classifications to select the sSFR threshold used to classify galaxies as SF or E.
We vary the threshold and check the consistency of the spectroscopic and photometric classifications.
{We optimized for a combination of the fraction of true E galaxies classified as E and the fraction of galaxies classified as E that were true E galaxies; these are the true positive rate and the positive predictive value.
The sSFR cut we adopt, $10^{-10}$ yr$^{-1}$, yields a true positive rate of 57\% and a positive predictive value of 73\%.
We use this value because deviations in either direction improve one of the metrics more slowly than they degrade the other metric.
For example, a more stringent cut of $10^{-11}$ yr$^{-1}$ yields a true positive rate of 24\% and a positive predictive value of 84\%: a $\approx$30\% decline for a $\approx$10\% improvement.
We use these metrics to choose the cut value because the sample consists mostly of SF galaxies, so mis-classification of SF galaxies as E galaxies is a greater problem than the reverse.
}
This selection process is why we use a higher sSFR threshold than some other works (e.g., \citealt{Werk:2012ug} used a cut of $10^{-11}$ yr$^{-1}$).

When we later associate \OVI\ absorption components with a galaxy, we will search for absorption components in a velocity window of $\pm 300$ km s$^{-1}$ centered on the galaxy's redshift.
If multiple galaxies could be associated with the same absorber, we associate the absorber to the galaxy with the smallest $\IP/\RCCc$.
In preparation for this association step, we remove galaxies from the sample if their redshift search windows overlap those of galaxies with smaller $\IP/\RCCc$.
This pre-processing step ensures that there is no double-counting of detected absorption systems or non-detections.
As part of this step, we classify galaxies as being isolated or non-isolated.
A galaxy is isolated if it has no removed neighbors with $\Mstar>10^{9.5}$ \Msun and non-isolated if it has at least one such neighbor.

We also check for cases where the chosen galaxy is within $\RCCc$\ of another galaxy with an overlapping redshift search window.
The purpose of this check is to find cases where absorption has been assigned to what may be a minor satellite of a much larger galaxy.
There are 18 chosen galaxies in the \CGMsq\ sample with another galaxy within $\RCCc$.
15 out of these 18 are more massive than any of the galaxies that are within $\RCCc$.
The remaining three galaxies have stellar masses that are within a factor of 3 of the most massive galaxy within $\RCCc$.

\subsubsection{Literature galaxy measurements}
\label{sec:data:galaxies:literature}
To increase our statistical power, we include several surveys from the literature in our sample.
These are: COS-Halos \citep{Werk:2012ug}, eCGM \citep{Johnson:2015tj}, dwarf galaxies from \citet{Johnson:2017tp}, the \citet{Keeney:2017wk} sample, the \citet{Keeney:2018wp} sample, the COS-LRG survey \citep{Chen:2018tu,Zahedy:2019vq}, the Red Dead Redemption (RDR) survey \citep{Berg:2019wq}, and a sightline from the QSAGE survey \citep{Bielby:2019vv}.
All of these surveys provide spectroscopic redshifts, which we adopt.
For galaxies from COS-Halos, eCGM, the \citet{Keeney:2018wp} sample, COS-LRG, and RDR, we estimate stellar masses and sSFRs by fitting publicly available photometric measurements using \texttt{CIGALE}.
We use the same set of photometric bands as for \CGMsq\ galaxies except for Gemini $g$ and $i$, which are not available for the literature galaxies.
Apart from this one difference in the available photometry, these masses and sSFRs are estimated using the same procedure as the \CGMsq\ galaxies.
The sSFRs are used to classify the galaxies as SF or E using the same cut as the \CGMsq\ galaxies, $10^{-10}$ yr$^{-1}$.

For three of the literature surveys, either adequately reduced public photometry covering optical to mid-infrared wavelengths is not available for some or all of the galaxies or additional measurements that are not part of a uniform public catalog are important for constraining the galaxy properties.
These include the following: dwarf galaxies from \citet{Johnson:2017tp} that are not in the eCGM sample; much of the \citet{Keeney:2017wk} sample, which includes resolved low redshift galaxies that are poorly measured by automated photometry pipelines; and the QSAGE galaxies.
We adopt the stellar masses published in these works.
For the \citet{Keeney:2017wk} and QSAGE galaxies, we also use their published star formation rates to calculate sSFRs and classify the galaxies as SF or E.
The classification is done using the same sSFR cut of $10^{-10}$ yr$^{-1}$ that was used for the rest of the sample.
\citet{Johnson:2017tp} do not provide star formation rates, but do state that all of their galaxies have blue colors and multiple detected emission lines.
We therefore assume these galaxies to be SF.
These surveys account for 26 of the 123 literature measurements in our sample.

Some of the surveys contain galaxies whose $\pm 300$ km s$^{-1}$ absorption search windows overlap.
As with \CGMsq, we remove galaxies whose search windows overlap those of a galaxy with smaller $\IP/\RCCc$.
In some cases, one of our literature sources will have used (and included in a table) a galaxy redshift from a different literature source that we also use.
This is the case, for example, with \citet{Johnson:2015tj}, which includes galaxies from COS-Halos.
To remove these duplicates, we match the complete galaxy sample (literature and \CGMsq) with itself using an angular tolerance of 4 arcseconds and a velocity tolerance of 75 km s$^{-1}$.
A galaxy that appears in the sample multiple times is ascribed to the first source in which it appears.

The distribution of the entire galaxy sample in the space of redshift, impact parameter, and stellar mass is shown in \autoref{fig:data:galaxies}.

\begin{deluxetable}{lccccc}
\tabletypesize{\footnotesize}
\tablecolumns{6}
\tablewidth{0pt}
\tablecaption{ Galaxy-absorber sub-sample sizes \label{tab:galaxy-sample-sizes}}
\tablehead{\nocolhead{} & \multicolumn{4}{c}{\IP\ Range (kpc)} & \nocolhead{} \\
\cline{2-5}
\colhead{Sub-sample} & \colhead{$0\rng100$} & \colhead{$100\rng200$} & \colhead{$200\rng300$} & \colhead{$300\rng400$} & \colhead{Total}}
\startdata
\CGMsq & 14 & 25 & 43 & 44 & 126 \\
Literature & 47 & 39 & 22 & 15 & 123\\
\hline
Star-forming & 46 & 51 & 52 & 48 & 197\\
Quiescent & 15 & 13 & 13 & 11 & 52\\
\hline
Total & 61 & 64 & 65 & 59 & 249\\
\enddata
\tablecomments{The number of galaxy-absorber pairs in the full sample and several sub-samples, broken down by galaxy impact parameter range. We split the sample by whether a galaxy-absorber pair is new or taken from the literature (\CGMsq\ vs. Literature) and, separately, by whether the galaxy in a galaxy-absorber pair is classified as star-forming or quiescent.}
\end{deluxetable}

\subsubsection{Sample Selection Biases}
\label{sec:data:galaxies:biases}
As is apparent in Figure \ref{fig:data:galaxies}, we are not sensitive to galaxies with \Mstar $<$ 10$^{9}$ M$_{\odot}$ at $0.5>z$.
This is obvious Malmquist bias, and we clearly are not offering a sample that is uniform in mass across all redshifts.
Thus, if there are significant trends in \OVI\ CGM content with redshift at $z<0.6$, our analysis may fail to find it and instead may possibly misinterpret it as a trend with mass.

The broad $z$ range is itself noteworthy.
$z=0 \rng 0.6$ covers $5.9 \times 10^9$ years, plenty of time for CGM properties to evolve.
Previous work on the evolution of the number of \OVI\ absorbers per unit redshift, \dNdz, over this redshift range has found that $\dNdz \propto (1+z)^{1.8 \pm 0.8}$ \citep{Danforth:2016uj}.
This redshift dependence is consistent with expectations for a population of absorbers with constant comoving cross section and comoving number density: $\dNdz \propto \frac{(1+z)^2}{E(z)}$, where $E(z)$ is the dimensionless Hubble parameter \citep{Bahcall:1969tt,Hogg:1999tj}.

\subsection{\OVI\ measurements}
\label{sec:data:ovi}
The second part of the dataset is a set of measurements of \OVI\ absorption in the spectra of QSOs.
The new \CGMsq\ measurements are made based on medium resolution far ultraviolet (FUV) spectra recorded with the \emph{Hubble Space Telescope} Cosmic Origins Spectrograph (\emph{HST} COS, \citealt{Green:2012vw}) using the G130M and G160M gratings.
We also include literature \OVI\ absorption measurements based on FUV spectra recorded with the \emph{HST} COS and the \emph{Far Ultraviolet Spectroscopic Explorer} (\emph{FUSE}, \citealt{Moos:2000vf,Moos:2002ux}).

\subsubsection{\CGMsq}
\label{sec:data:ovi:cgmsq}
The \emph{HST} COS spectra we use for \CGMsq\ were observed as part of the COS-Halos and COS-Dwarfs surveys (\citealt{Werk:2013uj,Bordoloi:2014uz}; Program IDs 11598 and 12248).
All 22 QSO sightlines were observed using both the G130M and G160M gratings, a combination which enables coverage of the \OVI\ doublet at  $z=0.10 \rng 0.72$.
The spectra have median signal-to-noise ratios (SNRs) of 5-12 per resolution element.
Within each spectrum, the SNR varies with wavelength by a factor of about two.
These SNRs allow $3\sigma$ detection of \OVI\ absorption down to column densities of $10^{13.5} \rng 10^{14.1}$ \cmmt.

For our analysis, we use the continuum normalized spectra produced by the COS-Halos and COS-Dwarfs teams.
Absorption features were initially identified through visual inspection by members of the \emph{Werk SQuAD}; this process is described in detail in \citetalias{Wilde:2021vr}.
We visually re-examined all identifications in wavelength ranges near \OVI\ wavelengths at the redshifts of \CGMsq\ galaxies and added or removed components in cases where the initial component structure did not provide a good fit to the absorption.

Based on these identifications, we measure \OVI{} column densities of detected absorption components using Voigt profile fitting.\footnote{Done with custom Voigt profile fitting software, checked against the package \texttt{veeper} (\url{https://github.com/jnburchett/veeper}).}
In spectral regions with no detected \OVI\ absorption, we calculate a 3$\sigma$ upper limit on the column density as estimated by the apparent optical depth method over a 100 km s$^{-1}$ velocity span, averaging the value of the limit over the span of a galaxy search window, $\pm300$ km s$^{-1}$.
That is, we assume that any un-detected absorption should fit within a 100 km s$^{-1}$ velocity span and average over the possible locations of this span within a search window.

Figure \ref{fig:specplot} shows \atomH\ and \OVI\ spectral regions at the redshifts of three \CGMsq\ galaxies.
When multiple absorption components are found in a galaxy's search window, their column densities are summed and this total column density is associated with the galaxy.
This is the case for the galaxies associated with the left and central columns of the Figure.
The spectra reflect the typical SNR range of the \CGMsq\ data and show two galaxies with \OVI\ detections and one galaxy with no detected \OVI.

\subsubsection{Literature \OVI\ measurements}
\label{sec:data:ovi:literature}
Most of the literature sources we use provide \OVI\ column densities or upper limits for each galaxy.
For these sources, we rescale their upper limits to also be 3$\sigma$ values over 100 km s$^{-1}$ spans.
Due to the differing goals of each survey, the spectra they used to measure \OVI\ have a range of SNRs (6-17) and yield a range of limiting \OVI\ column densities.
The broad range of limiting column densities requires us to use analysis methods that consistently extract information from upper limits that are greater than some detections.
This requirement motivates how we analyze $\NOVI(\Mstar, \IP)$ around star-forming galaxies in \S \ref{sec:analysis} and forces us to use a survival function-based estimator to calculate median column densities.

\citet{Johnson:2017tp} only provide \OVI\ column density measurements for two of the galaxies in their sample.
For consistency, we re-measure column densities and upper limits at the redshifts of all of the galaxies in their sample using the same Voigt profile fitting procedure we used for \CGMsq.
The measurements are done based on spectra downloaded from the Hubble Spectroscopic Legacy Archive \citep{Peeples:2017wd} and continuum-normalized using the \texttt{linetools}\footnote{\url{https://doi.org/10.5281/zenodo.168270}} continuum normalization interface.

\subsection{Associating galaxies with \OVI\ absorption}
\label{sec:data:synthesis}
We choose to ascribe each absorber to a single galaxy.
In situations where absorption is within the search windows ($\pm 300$ km s$^{-1}$) of multiple galaxies, we ascribe it to the galaxy with the smallest halo radius-normalized impact parameter $\IP/\RCCc$.
Figure \ref{fig:NOVI-vs-gal-params-overview} shows the \OVI\ column densities as a function of impact parameter and stellar mass for our final sample with this
absorber-to-galaxy assignment method.
This association criterion does also mean that some \OVI\ absorption components are excluded from the survey as a result of being associated solely with an excluded galaxy.
This exclusion happens when two or more galaxies have overlapping search windows where an \OVI\ absorption component is present outside the window of the galaxy with the smallest $\IP/\RCCc$.
For example, suppose there is a galaxy at $v=0$ km s$^{-1}$, another galaxy is at $200$ km s$^{-1}$, and an absorption component is at 400 km s$^{-1}$.
If the galaxy at $0$ km s$^{-1}$ has a smaller $\IP/\RCCc$ than the the galaxy at $200$ km s$^{-1}$, this absorption component would be excluded.
If no absorption is detected in a galaxy's search window, the galaxy is associated with an upper limit measured at the center of its search window.
This limit is already an average over the search window by construction (see \S \ref{sec:data:ovi:cgmsq}).

\begin{figure*}
    \centering
    \includegraphics[width=\linewidth]{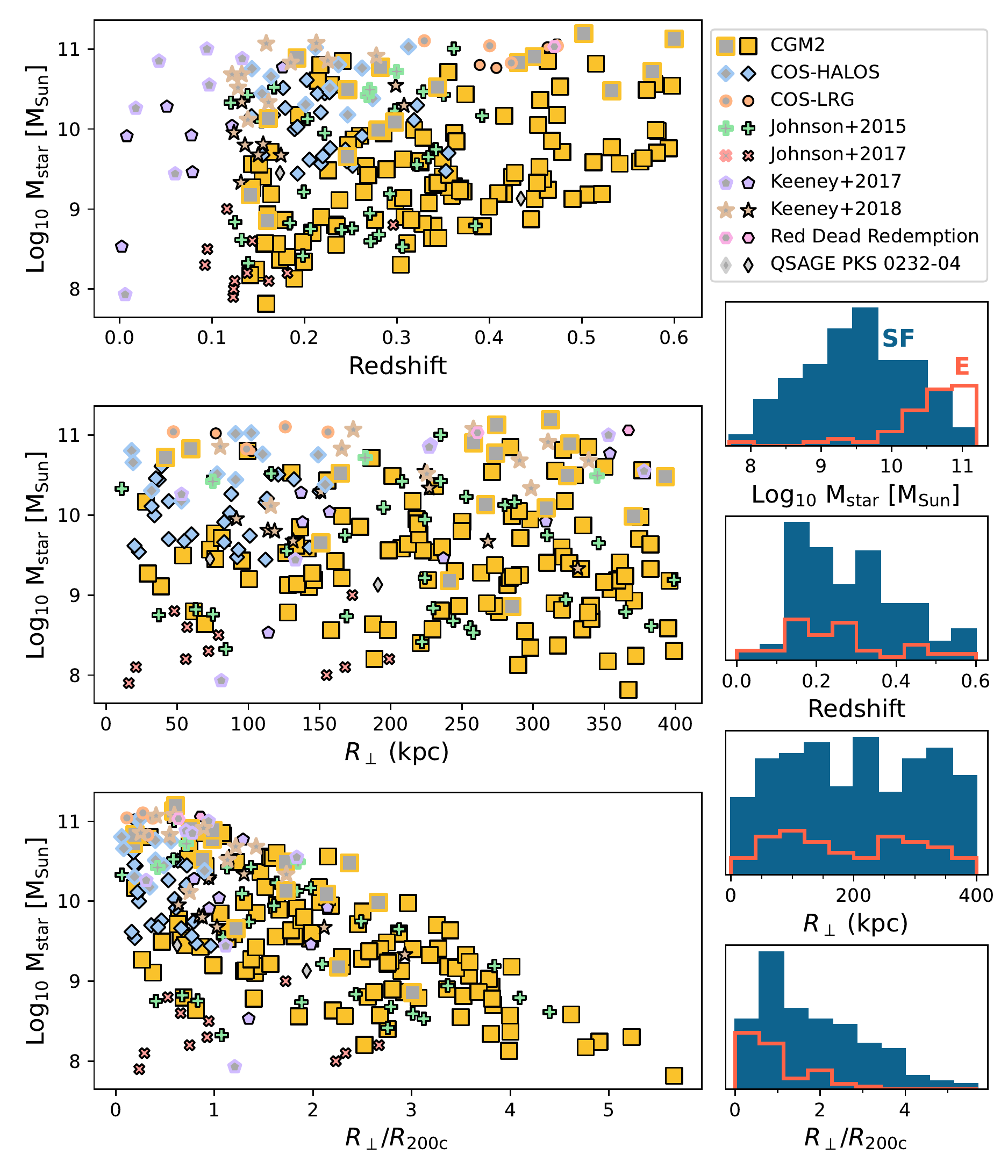}
    \caption{
    Redshifts, impact parameters, $\RCCc$-normalized impact parameters, and stellar masses of galaxies in the sample.
    The sample includes galaxies from multiple source surveys; the source of each galaxy is indicated in the three left panels by the shape and color of the galaxy's datapoint.
    {
    Star-forming (SF) galaxy datapoints have black outlines and are filled in with the color corresponding to their survey; quiescent (E) galaxy datapoints are outlined in the color corresponding to their survey and are filled in gray.
    The distributions of SF and E galaxies in the four parameters listed above are shown in the four panels on the right.}
    References to the surveys are listed in Table \ref{tab:galaxy-samples}.
    Galaxies assigned to the \CGMsq\ survey have not been studied in earlier surveys.
}
    \label{fig:data:galaxies}
\end{figure*}

\begin{figure*}
  \includegraphics[width=\linewidth]{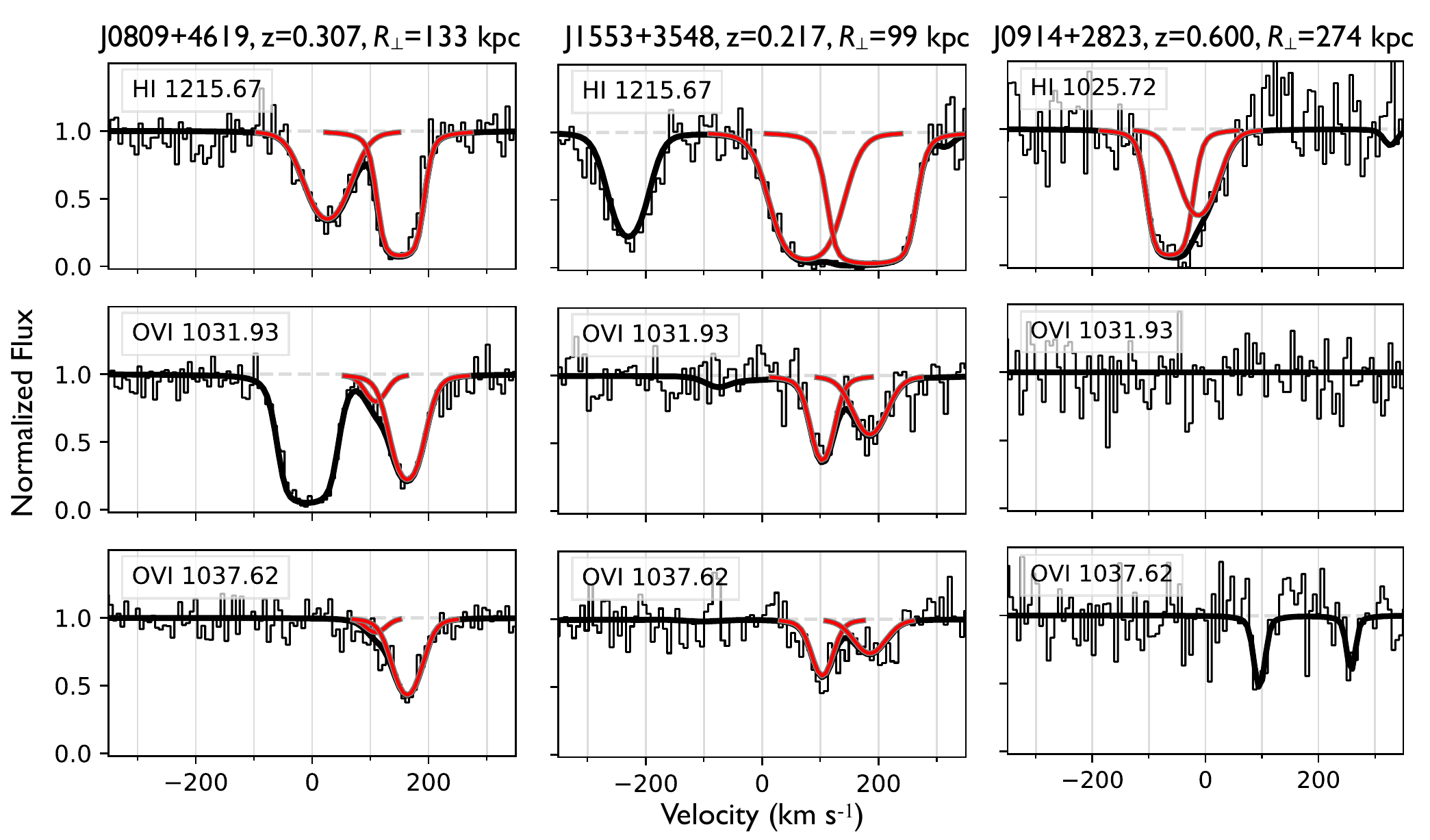}
  \caption{
  \atomH{} and \OVI{} absorption near the redshifts of three \CGMsq{} galaxies.
  Each column of panels is associated with a different galaxy.
  The black line is part of a fit to the entire COS spectrum.
  The red lines are components of the fit that are due to the transition indicated by the panels' labels.
  The spectrum shown in the left column has a high signal-to-noise ratio relative to the rest of our sample; the signal-to-noise ratios of the spectra shown in the central and right columns are more typical.
  The left and middle columns show examples of \OVI\ detections while the right column shows a case where \OVI\ is not detected.
  }
  \label{fig:specplot}
\end{figure*}

\begin{figure*}
  \includegraphics[width=\linewidth]{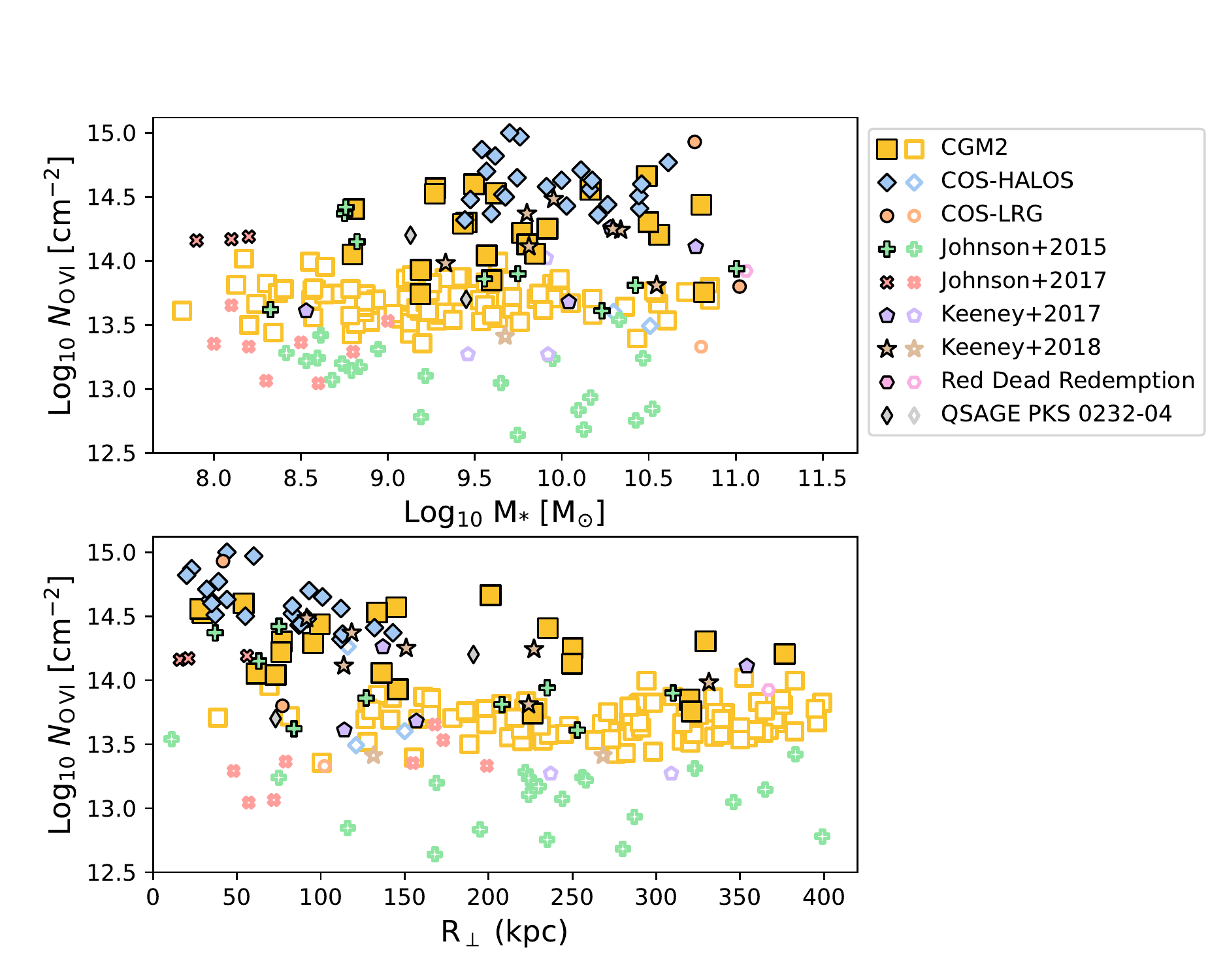}
  \caption{
  \OVI\ column densities as a function of the stellar masses and impact parameters of {star-forming} absorber host galaxies.
  Filled data points outlined in black are \OVI\ detections; hollow data points outlined in color are 3$\sigma$ upper limits associated with \OVI\ non-detections.
  The shape and color (fill color for a detection, edge color for a non-detection) of a data point indicate the source survey for a host galaxy.
  References to the surveys are listed in Table \ref{tab:galaxy-samples}.
  Host galaxies assigned to the \CGMsq{} survey have not been studied in earlier surveys.
}
  \label{fig:NOVI-vs-gal-params-overview}
\end{figure*}

We make an additional correction to exclude \CGMsq\ galaxies where \OVI\ line identifications or limits are made uncertain by the presence of detector gaps, geocoronal \specnotation{O}{I} emission, or strong absorption from gas at other redshifts.
Limits were considered uncertain when some combinations of these three factors affected the same substantial velocity range for both \OVI\ lines.
For example, strong unrelated absorption covering $0\rng 300$ km s$^{-1}$ relative to the galaxy redshift would lead us to exclude that galaxy.
Detections were considered uncertain when the factors rendered one of the \OVI\ lines uninformative.
A detection where the \OVI\ component of the fit is the dominant contributor to the total profile at its location would be included; a detection where the \OVI\ absorption is weaker than that of the blended line or lines would not.
5 out of 131 \CGMsq\ galaxies were removed, 2 of which were non-detections and 3 of which were detections according to the initial assessment.
The results of the analyses we describe later in this work do not significantly change when we include the measurements associated with these 5 galaxies.
These 5 galaxies are not shown in Figures \ref{fig:data:galaxies} or \ref{fig:NOVI-vs-gal-params-overview}.

In total, the sample consists of 249 galaxy-absorber pairs.
The number of pairs in different sub-samples---SF or E, new or taken from the literature---and at different impact parameter ranges is listed in \autoref{tab:galaxy-sample-sizes}.
Data for each pair is given in \autoref{tab:galaxies-OVI}.
The \OVI\ column density \NOVI\ is shown as a function of galaxy properties in \autoref{fig:NOVI-vs-gal-params-overview}.

From \autoref{fig:NOVI-vs-gal-params-overview}, it is clear that there are trends in \NOVI\ as a function of both galaxy stellar mass \Mstar\ and \IP.
The upper envelope of \NOVI\ as a function of \Mstar\ increases from $10^8$ to $10^{9.5}$ \Msun\ and decreases above $10^{10.5}$ \Msun.
The fraction of \OVI\ upper limits or sub-$10^{14}$ \cmmt\ detections increases with increasing impact parameter.
We explore this bivariate trend for the star-forming galaxies in the sample in \S \ref{sec:analysis}.

Because there are too few quiescent galaxies in our samples with $\Mstar\lesssim 10^{10}$ \Msun, we do not explore the broad mass dependence of \OVI\ absorption around quiescent galaxies.
An initial comparison between \NOVI\ at fixed \Mstar\ and \IP\ around star-forming and quiescent galaxies was inconclusive.
Whether or not there is a significant difference depends on the treatment of the dependence of \NOVI\ on galaxy mass.
If we assume the mass dependence is locally (i.e., over a bin in mass) constant, there is difference.
If we assume the mass dependence is linear in $\logten \Mstar$-$\logten \NOVI$ space, there is no difference within the uncertainties, but the uncertainties are greater.
Determining which scenario is true will require a careful and specialized analysis that we leave to future work.

\begin{deluxetable*}{llccccccccc}
\tabletypesize{\footnotesize}
\tablecolumns{11}
\tablewidth{0pt}
\tablecaption{ Galaxy-absorber pairs \label{tab:galaxies-OVI}}
\tablehead{
\colhead{Source survey} & \colhead{QSO} & \colhead{R.A.} & \colhead{Decl.} &
\colhead{Redshift} & \colhead{\IP} & \colhead{$\logten$\Mstar} & \colhead{SF/E} &
\colhead{\RCCc} & \colhead{$\logten$\NOVI} & \colhead{Limit flag}
\vspace{-0.06in}\\
\nocolhead{} & \nocolhead{} &
\colhead{(J2000)} & \colhead{(J2000)} & \nocolhead{} & \colhead{(kpc)} & \colhead{[\Msun]} & \nocolhead{} & \colhead{(kpc)} & \colhead{[\cmmt]} & \nocolhead{}
}
\startdata
Keeney+2018 & HE 0153-4520 & 28.8038 & -45.1095 & 0.2252 & 80 & 10.9 & E & 316 & 14.21 &  \\
eCGM & HE 0226-4110 & 37.0426 & -40.9300 & 0.1248 & 230 & 8.8 & SF & 95 & 13.17 & $<$ \\
eCGM & HE 0226-4110 & 37.0865 & -40.9471 & 0.1992 & 224 & 9.2 & SF & 107 & 13.1 & $<$ \\
eCGM & HE 0226-4110 & 37.0607 & -40.9563 & 0.2065 & 37 & 8.8 & SF & 91 & 14.37 &  \\
eCGM & HE 0226-4110 & 37.0679 & -40.9576 & 0.2678 & 75 & 10.4 & E & 176 & 13.14 & $<$ \\
eCGM & HE 0226-4110 & 37.0933 & -40.9497 & 0.2706 & 345 & 10.5 & E & 187 & 13.1 & $<$ \\
eCGM & HE 0226-4110 & 37.0472 & -40.9435 & 0.2804 & 244 & 8.7 & SF & 87 & 13.07 & $<$ \\
eCGM & HE 0226-4110 & 37.0390 & -40.9624 & 0.3341 & 346 & 9.6 & SF & 120 & 13.05 & $<$ \\
eCGM & HE 0226-4110 & 37.0794 & -40.9668 & 0.3416 & 310 & 9.7 & SF & 125 & 13.9 &  \\
COS-Halos & J0226+0015 & 36.5541 & +0.2581 & 0.2274 & 78 & 10.5 & E & 193 & 13.32 & $<$ \\
\enddata
\tablecomments{
Data on galaxy-absorber pairs in the sample. The impact parameter \IP\ and the virial radius \RCCc\ are given in physical, not co-moving, kiloparsecs. References to the different source surveys are given in \autoref{tab:galaxy-samples}.\\
(This table is available in its entirety in machine-readable form in the published text.)}
\end{deluxetable*}

\section{The dependence of CGM \OVI\ content on stellar mass and impact parameter in star-forming galaxies}
\label{sec:analysis}

\begin{figure*}
    \centering
    \includegraphics{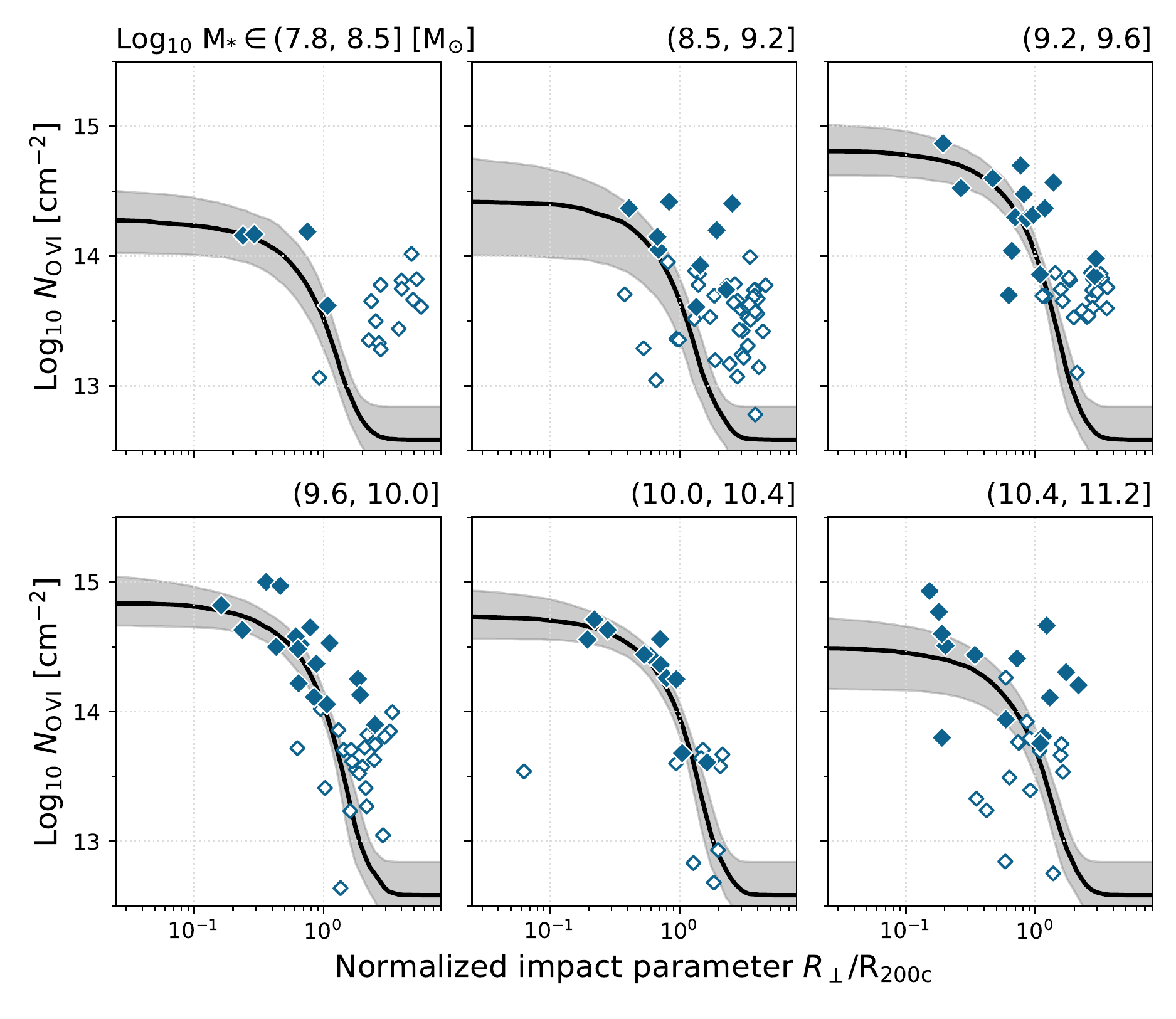}
    \caption{
    Measured and model \OVI{} column densities as a function of host galaxy stellar mass and impact parameter to the quasar sightline.
    The median model $\NOVI(\IP)$ profile for each mass bin is shown as a black line.
    Model uncertainty is represented by the gray region about each black line, which shows the 16th and 84th percentiles of $\NOVI(\IP)$.
    The model is fit to \OVI{} associated with star-forming galaxies (blue diamonds).
    Filled diamonds represent \OVI{} detections while hollow diamonds  are 3$\sigma$ upper limits on non-detections.
    The model is the sum of an $\IP$-dependent profile and a constant background and is described in more detail in the first part of Section \ref{sec:analysis}.
    Controlling for impact parameter helps isolate the mass dependence of \OVI{} column density on galaxy stellar mass.
    }
    \label{fig:expfit}
\end{figure*}

This section is concerned with how (1) the column density and (2) the spatial extent of \OVI\ in a SF galaxy's CGM depends on the galaxy's stellar mass.
We make the assumption that the \OVI\ that we have associated with a galaxy A is part of galaxy A's CGM, and not part of the CGM of galaxy B that happens to be near galaxy A because of galaxy clustering.
The validity of our assumption depends on how complete the input galaxy surveys are and how correct associations between detected galaxies and \OVI\ absorbers are.
The effect of contamination from neighboring galaxy halos will be strongest when the assumed host galaxy intrinsically has low \NOVI.
This may be expected for low mass galaxies and for massive galaxies, groups, and clusters where the halo temperature collisionally ionizes oxygen more than 5 times.
We discuss how appropriate the assumption of a single CGM host galaxy is for the low mass ($\Mstar \lesssim 10^{9}$ \Msun) part of our sample in \S \ref{sec:discussion:low-mass-association}.

A meaningful comparison of CGM behavior between halos of different masses and sizes requires a standardized and well-defined summary quantity to compare.
This requirement drives our choice of analysis strategy---training a parametric model for \NOVI\ as a function of \Mstar\ and \IP and using this model to compute summaries.
A strategy based on parametric modeling allows us to correct for the non-random spatial distribution of the galaxy sample.
As a side benefit, it also provides a clear way of quantifying CGM extent and its dependence on \Mstar.

We cannot simply use the sample to calculate summaries because the dataset is not a geometrically uniform sampling of the CGM out to our maximum investigated \IP\ of 400 kpc.
Furthermore, the non-uniformity is correlated with galaxy mass.
These issues arise because we are using a collection of galaxy surveys with different selection criteria in impact parameter and galaxy mass.
For example, the COS-Halos galaxies were selected to have $\IP<150$ kpc and the \citet{Johnson:2017tp} galaxies were selected to be within three virial radii of a sightline (corresponding to $\IP<260$ kpc for the most massive galaxy in their sample) and to oversample galaxies within one virial radius of the sightline.
A CGM summary quantity, such as a covering fraction within some radius, a mean \NOVI, or an \OVI\ mass $M_{\OVI}$, computed using this dataset without some form of correction would be biased towards the properties of the oversampled inner CGM.

Using the data to train a parametric model and then using the parametric model to calculate summary quantities resolves this issue.
The density of samples no longer determines the weight of an \IP\ range in an integral. Instead, it determines how the uncertainty of the parametric model depends on \IP.
This uncertainty can then be propagated to the summary quantities.
Using a parametric model instead of using the un-weighted data actually makes non-geometrically uniform sampling an advantage.
If our sample of 55 sub-$10^9$ \Mstar\ dwarf galaxies were geometrically uniform, 90\% of samples would contain 5 or fewer galaxies within \RCCc\ of the sightline.
The actual sample contains 12 and so provides more information about \OVI\ in the CGM of dwarf galaxies.

\subsection{An empirical model for \OVI\ around star-forming galaxies}
\label{sec:analysis:sf-gal-model}

The model is trained using the \Mstar, \IP, and $\log_{10}\NOVI$\ values of the SF galaxy-absorber sample.
As output, the model describes the typical $\log_{10}\NOVI$\ and the expected intrinsic $\log_{10}\NOVI$\ scatter as a function of \Mstar\ and \IP.
The \NOVI\ contribution of a galaxy is assumed to be a declining exponential function of \IP\ raised to a power ($\IP^\gamma$).
This functional form includes a log-linear dependence as a special case, $\gamma=1$, but allows more variation of the shape of the \IP\ dependence.
\NOVI\ is assumed to be piecewise-constant in \Mstar.
In effect, we are fitting exponentials in $\IP^\gamma$\ to collections of galaxies binned according to their \Mstar.

We split the galaxies into six \Mstar\ bins with bin edges $\logten (\Mstar/\Msun)=$7.8, 8.5, 9.2, 9.6, 10, 10.4, and 11.2.
The bins are chosen to balance three goals: (1) narrow extent in $\Mstar$; (2) large sample size within \RCCc\ for each bin; and (3) large sample size over all radii for each bin.
Within each bin, the dependence of \NOVI\ on impact parameter is described by the superposition of a constant baseline level and the exponential profile in $\IP^\gamma$:
\begin{equation}
\begin{split}
    \logten \NOVI(\Mstar, \IP) =& \\\logten \biggl(
    10^{\Nzero}& e^{-\left(\frac{\IP}{\scalelen \RCCc}\right)^\gamma}
    + 10^{\Nbase} \biggr)
\end{split}
\end{equation}
\label{eqn:expon-prof}
The width and shape of the profile are determined by a length scale parameter $\scalelen$ and a power $\gamma$.
The value of the profile at $\IP=0$ is determined by the parameter $\Nzero$.
The constant baseline level is determined by the parameter $\Nbase$.
The baseline reflects the fact that the expected column density along a random redshift interval is small, but non-zero.
\citet{Nelson:2018wy} estimate a mean non-halo \NOVI\  value of $10^{12.5}$ \cmmt\ using results from \citet{Danforth:2016uj}.
This non-halo \OVI\ may come from the halos of other, undetected galaxies or from the intergalactic medium (IGM).

This superposition describes the \emph{typical} value of \NOVI\ at a given host galaxy stellar mass and impact parameter.
Observed values are scattered about this typical value.
Over a small range in \Mstar\ and \IP, one can find detections with $\NOVI \geq 10^{14}$ \cmmt\ and upper limits with $\NOVI < 10^{13.2}$ \cmmt.
A typical $\log_{10}$\NOVI\ uncertainty for our sample is $0.1 \rng 0.2$ dex, suggesting that there is a source of scatter besides measurement uncertainty.
We assume this scatter is mostly physical.
It may reflect differences in galaxy parameters that we do not control for (e.g., inclination, sSFR within the SF and E classes) or random variation due to inhomogeneities in the CGM.

The deviations are assumed to follow a two-sided exponential, or Laplace, distribution:
\begin{equation}
\label{eqn:laplace-dist}
    p(\logten N_i) = \frac{1}{2a} \exp\left(-\frac{1}{a} \vert \logten N_i - \log_{10}\NOVI(M_i,R_{\perp,i}) \vert \right).
\end{equation}
The parameter $a$ describes the amplitude of this intrinsic scatter.
We have chosen a Laplace distribution because it is less sensitive to outliers than a Gaussian distribution.
We expect that our results would change little if we instead used a different heavy-tailed and symmetric distribution.

\autoref{eqn:laplace-dist} describes measurements where $N_i$ has been detected.
We extend the equation to the case of upper limits by marginalizing over $N_i$.
We treat an upper limit as a statement that the column density is less than the limit with probability $p_s$ and greater than the limit with probability $1-p_s$.
$p_s$ is set by the significance of the limit (e.g., 99.73\% for $3\sigma$).
The posterior probability of an upper limit given a model for the typical value and scatter is then the sum of two products: (1) the probability of being below the limit according to \autoref{eqn:laplace-dist} and the limit's significance and (2) the probability of being above the limit according to \autoref{eqn:laplace-dist} and the limit's significance:
\begin{equation}
\begin{split}
    p(\text{Limit }&N_i) = p_s \int_{-\infty}^{\logten N} p(\logten N') \dd \logten N'\\
    &+ (1-p_s) \int_{\logten N_i}^{+\infty} p(\logten N') \dd \logten N',
\end{split}
\end{equation}
where $p(\logten N')$ is given by \autoref{eqn:laplace-dist}.

In total, there are up to five parameters per mass bin: $\scalelen$, $\Nzero$, $\Nbase$, $a$, and $\gamma$.
We find that when all five parameters are fit independently for each bin, the $68\%$ credible intervals for the $\scalelen$, $\Nbase$, $a$, and $\gamma$ parameters agree from mass bin to mass bin.
That is, the available data do not require these parameters to depend on \Mstar\ over the mass range we are working in.
For the results shown in this work, we assume these four parameters are shared by the different mass bins while each mass bin has its own zero-point value $\Nzero$.
Since the parameters are consistent across mass bins even when not assumed to be shared, this assumption does not significantly affect our results while reducing the number of parameters in the problem.

We infer parameters and their uncertainties by generating samples from the model's posterior probability distribution.
Sampling is done using the \texttt{PyMC3} modeling framework \citep{Salvatier:2016ul}.
We use these samples to calculate medians and quantile-based credible intervals for the parameters.
These values are reported in \autoref{tab:exponential-model-param-results}.
The median model and its uncertainty as seen in terms of the observables is shown in \autoref{fig:expfit}.
When we later calculate quantities that are derived from the model, we propagate uncertainties using samples.
For example, to estimate the uncertainty on the total mass of \OVI\ within a virial radius of a galaxy (see \S \ref{sec:analysis:sf-mass-trend}), we first calculate the total mass according to each set of sampled parameter values, then use that collection of total mass estimates to compute uncertainties.

\begin{deluxetable}{lc}
 \tabletypesize{\footnotesize}
 \tablewidth{\linewidth}
\tablecaption{
Estimated \NOVI(\Mstar, \IP)\ model parameters
\label{tab:exponential-model-param-results}}
\tablecolumns{2}
\tablehead{\colhead{Parameter} & \colhead{Estimate}}
\startdata
Scale $\scalelen$ & $0.7^{+0.2}_{-0.3}$ \\[3pt]
Norm $\Nzero$, $7.8 \rng 8.5$ &  $14.3^{+0.2}_{-0.3}$\\[3pt]
Norm $\Nzero$, $8.5 \rng 9.2$ &  $14.4^{+0.3}_{-0.5}$\\[3pt]
Norm $\Nzero$, $9.2 \rng 9.6$ &  $14.8^{+0.2}_{-0.2}$\\[3pt]
Norm $\Nzero$, $9.6 \rng 10$ &  $14.8^{+0.3}_{-0.1}$\\[3pt]
Norm $\Nzero$, $10 \rng 10.4$ & $14.7^{+0.2}_{-0.1}$\\[3pt]
Norm $\Nzero$, $10.4 \rng 11.2$ & $14.5^{+0.2}_{-0.3}$\\[3pt]
Scatter amplitude $a$ & $0.5^{+0.1}_{-0.1}$\\[3pt]
Baseline $\Nbase$ & $12.6^{+0.2}_{-0.6}$ \\[3pt]
Power $\gamma$ & $1.6^{+0.9}_{-0.6}$\\[3pt]
\enddata
\tablecomments{Parameter estimates for the \NOVI\ profile model defined in \autoref{sec:analysis:sf-gal-model}.
The point estimate is the median of the distribution.
Uncertainties are calculated from the 16th and 84th percentiles of the distribution.
The scale, baseline, scatter amplitude, and power parameters are shared across \Mstar\ bins, but each \Mstar\ bin has its own norm parameter.
The range given in the name of a norm row is the $\log_{10}$ of the range \Mstar\ range in \Msun\ units; for example, "Norm $\Nzero$, 7.8-8.5" is the norm for the mass bin covering $\Mstar=10^{7.8} \rng 10^{8.5}$ \Msun.}
\end{deluxetable}

To check for unmodeled or inadequately modeled trends between \NOVI\ and \Mstar, \IP, and $z$, we examine residuals from the model as a function of these parameters.
We split the full sample into sub-samples defined using these parameters and calculate the median residual in each sub-sample.
If there are significant univariate trends in the typical \OVI\ column density that our model does not reflect, the median residual should be non-zero.
We use the Kaplan-Meier survival function estimator as implemented in \texttt{lifelines}\footnote{\url{https://doi.org/10.5281/zenodo.805993}} to calculate these medians.
The use of this estimator is necessary because of the presence of \NOVI\ limits.
Uncertainties on these medians are estimated using bootstrapping.

\Mstar\ sub-samples are defined by splitting each mass bin at its logarithmic midpoint.
For example, the $10^{9.6} \rng 10^{10}$ \Msun\ mass bin is split into $10^{9.6} \rng 10^{9.8}$ and $10^{9.8} \rng 10^{10}$ \Msun\ sub-samples).
We create three \IP\ sub-samples with boundary values $\IP/\RCCc=0, 0.5, 1, 5.7$ and three $z$ sub-samples with boundary values $z=0, 0.2, 0.4, 0.6$.
The largest \IP\ boundary value is the maximum $\IP/\RCCc$ of the full sample.

The medians of the \Mstar\ and $z$ sub-sample residuals are all consistent with 0.
The data do not show an obvious need for a more complicated dependence on \Mstar\ beyond what is included in the model or for an explicit dependence on $z$.
The medians of the two inner \IP\ sub-sample residuals are also consistent with 0, but the median of the outer \IP\ sub-sample residuals is not.
In \autoref{fig:expfit}, there are indeed some observations outside of \RCCc\ that are well above the model prediction.
Examining the properties of such outliers in the \CGMsq\ part of the sample, we find that many of them have neighbors with $\Mstar>10^{9.5}$ \Msun.
This tendency suggests that these galaxies' greater-than-typical \NOVI\ is due to the environmental effect noted by  \citet{Johnson:2015tj}.
We discuss this possibility further in \S \ref{sec:discussion:environment}.

{
To test for the possibility of systematic offsets between source surveys, we re-fit the model while excluding different sub-samples of the dataset.
We consider three cases: excluding COS-Halos, excluding eCGM, and excluding surveys that contribute fewer than 30 galaxy-absorber pairs each (i.e., using only \CGMsq, COS-Halos, and eCGM).
For all parameters but one, the 16-to-84\% credible intervals of the sub-sample solutions and the full-sample solution overlap.
The exception is the norm $\Nzero$\ of the $\Mstar=10^{8.5}\rng10^{9.2}$ \Msun\ bin.
Removing the eCGM galaxy-absorber pairs from the sample shifts the upper bound of this parameter's credible interval to 13.6, 0.3 dex lower than the lower bound of the full sample credible interval.
This shift is the result of 3 of the 4 \OVI\ detections at $\RCCc<1$ in this mass bin being drawn from the eCGM sample, a coincidence that reflects the relatively small number of dwarf galaxy-absorber pairs at low impact parameters.}

{
The main effect of excluding sub-samples is the expansion of the credible intervals.
This expansion is not always uniform and symmetric across the exclusion cases.
Using only the three largest surveys removes all but one of the sub-$\RCCc$\ measurements in the lowest-mass bin, dramatically increasing the uncertainty on that bin's normalization.
Excluding COS-Halos removes a large fraction of the $\RCCc<0.5$ measurements at $\Mstar>10^{9.2}$ \Msun.
This removal expands the $\Nzero$\ credible intervals for the higher mass bins by shifting the lower bounds of the intervals to smaller values.
To summarize these results, the tests do not detect systematic differences between the input surveys once galaxy mass and impact parameter have been controlled for.
}

\subsection{Average \OVI\ content of the SF galaxy CGM as a function of galaxy stellar mass}
\label{sec:analysis:sf-mass-trend}

\begin{figure}
    \centering
    \includegraphics[width=\linewidth]{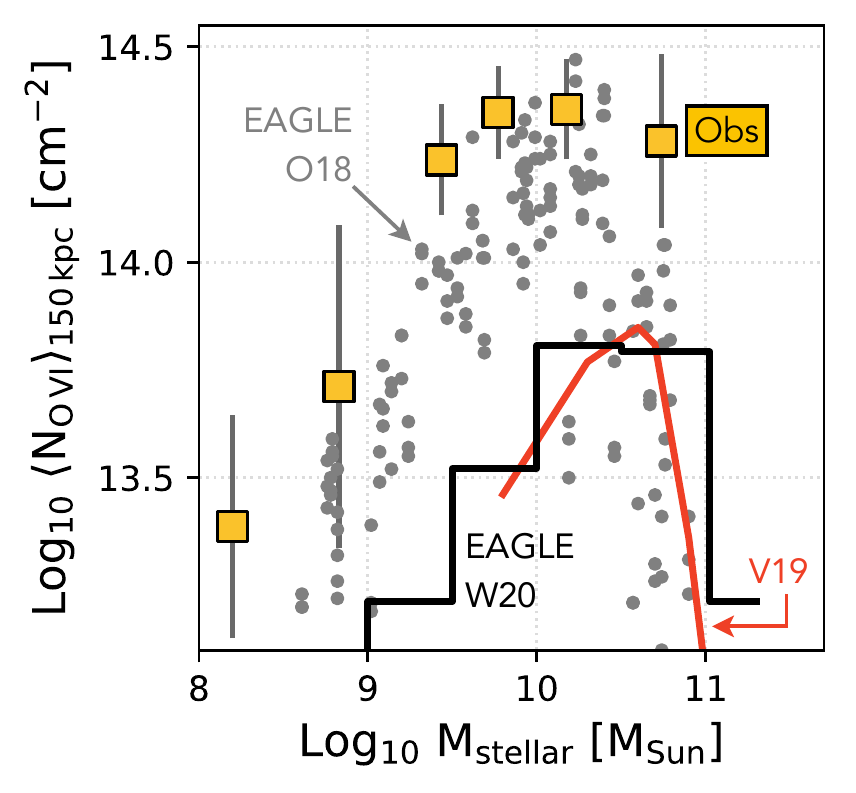}
    \caption{
    Average column densities of \OVI\ within 150 physical kpc (\ApNOVI{\text{150 kpc}}) of galaxies with different stellar masses.
    We show \ApNOVI{\text{150 kpc}} derived from: our measurements of star-forming galaxies (yellow squares; labeled Obs); \emph{EAGLE} zoom simulations from \citet{Oppenheimer:2018vn} (gray circles; EAGLE O18); our implementation of the \citet{Voit:2019tv} precipitation-limited halo model (red line; V19); and the \emph{EAGLE} cosmological simulation from \citet{Wijers:2020vr} (solid black line; EAGLE W20).
    The \emph{EAGLE} zoom simulations provide the closest match to our measurements, but still tend to have lower \ApNOVI{\text{150 kpc}}.
    }
    \label{fig:mstar-novi-150}
\end{figure}

\begin{figure*}
    \centering
    \includegraphics[width=\linewidth]{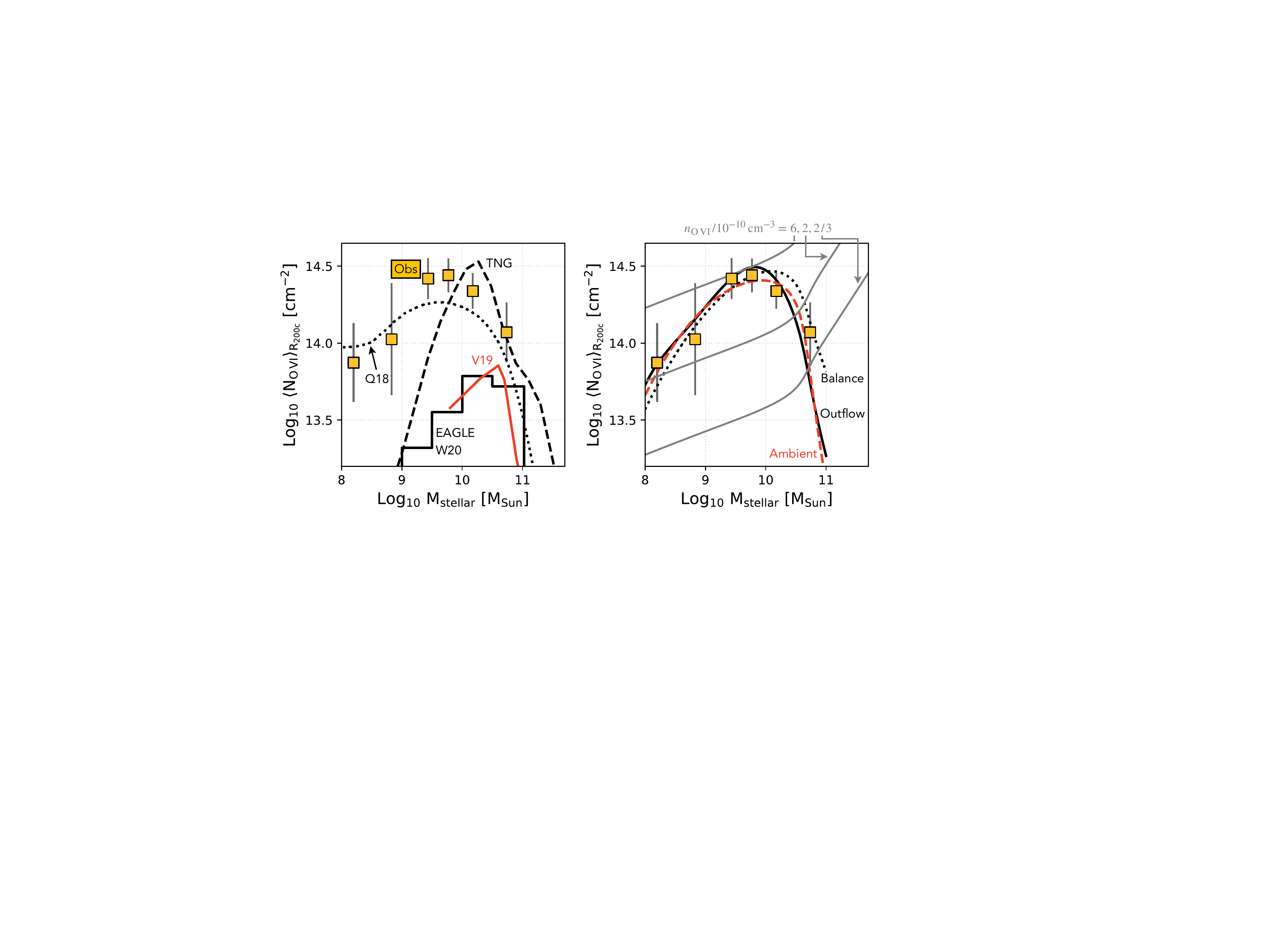}
    \caption{
    Average column densities of \OVI\ within a virial radius (\ApNOVI{\RCCc}) of galaxies with different stellar masses.
    We show \ApNOVI{\RCCc} derived from our measurements of star-forming galaxies (yellow squares; labeled Obs) in both panels of the figure.
    In the left panel, we show \ApNOVI{\RCCc} derived from: the TNG-100 cosmological simulation from \citet{Nelson:2018wy} (dashed line; TNG); the \citet{Qu:2018va} model (dotted line; Q18); our implementation of the \citet{Voit:2019tv} precipitation-limited halo model (solid red line; V19); and the \emph{EAGLE} cosmological simulation from \citet{Wijers:2020vr} (solid black line; EAGLE W20).
    In the right panel, we show \ApNOVI{\RCCc} derived from: an ambient phase scaling relation (dashed red line; Ambient); a star formation-driven hot outflow scaling relation (solid black line; Outflow); and a star formation and feedback-balancing inflow scaling relation (dotted black line; Balance).
    We also show lines corresponding to different constant $n_{\OVI}$ values (light gray).
    The scaling relations are defined in Equations \ref{eqn:ambient-scaling}, \ref{eqn:outflow-scaling}, and \ref{eqn:inflow-scaling}.
    }
    \label{fig:mstar-novi-r200c}
\end{figure*}

\begin{figure*}
    \centering
    \includegraphics[width=\linewidth]{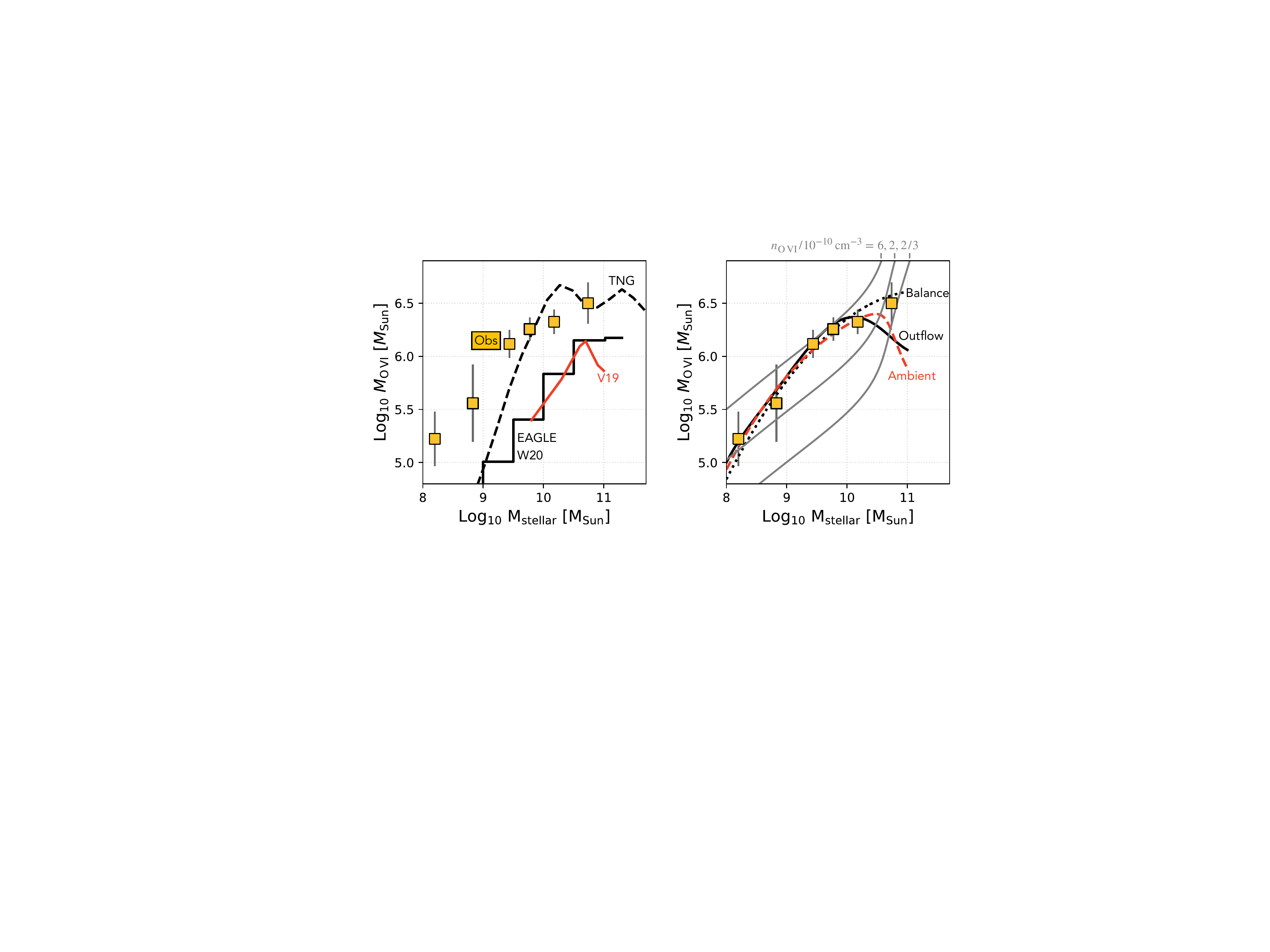}
    \caption{
    Masses of \OVI\ within a virial radius ($M_{\OVI}$) of galaxies with different stellar masses.
    We show $M_{\OVI}$ derived from our measurements of star-forming galaxies (squares; labeled Obs) in both panels of the figure.
    In the left panel, we show $M_{\OVI}$ derived from: the TNG-100 cosmological simulation from \citet{Nelson:2018wy} (dashed line; TNG); our implementation of the \citet{Voit:2019tv} precipitation-limited halo model (solid red line); and the \emph{EAGLE} cosmological simulation from \citet{Wijers:2020vr} (solid black line; EAGLE W20).
    In the right panel, we show $M_{\OVI}$ derived from: an ambient phase scaling relation (dashed red line; Ambient); a star formation-driven hot outflow scaling relation (solid black line; Outflow); and a star formation and feedback-balancing inflow scaling relation (dotted black line; Balance).
    We also show lines corresponding to different constant $n_{\OVI}$ values (light gray).
    We also show lines corresponding to different constant $n_{\OVI}$ values (light gray).
    The scaling relations are defined in Equations \ref{eqn:ambient-scaling}, \ref{eqn:outflow-scaling}, and \ref{eqn:inflow-scaling}.
}
    \label{fig:mstar-movi}
\end{figure*}

In this section, we examine how the \OVI\ content of a star-forming galaxy's CGM depends on the galaxy's stellar mass.
We use our parametric model for \NOVI(\Mstar, \IP)\ to calculate summary statistics for the total \OVI\ content of the star-forming galaxy CGM at different galaxy masses and compare these statistics with values from simulations, models, and scaling relations.
From these calculations and comparisons, we draw three main conclusions:
\begin{enumerate}
    \item The average \OVI\ column density within \RCCc\ of SF galaxies is greatest for galaxies with $\Mstar=10^{9.2}$ to $10^{10}$ \Msun.
    This is also the mass range where the equivalent average volume density of \OVI\ within \RCCc\ is greatest.
    {
    The average \OVI\ column density within 150 kpc of SF galaxies is greatest above $\Mstar=10^{9.2}$ \Msun\ but does not have a distinct peak mass.
    These results are shown in Figures \ref{fig:mstar-novi-150} and \ref{fig:mstar-novi-r200c}.}
    \item We {construct} three scaling relations that adequately describe the mass dependence of the \OVI\ content of a star-forming galaxy's CGM.
    In the first of these relations, \OVI\ is found in a collisionally ionized ambient phase with temperatures distributed around the halo virial temperature.
    In the other two relations, the amount of \OVI\ in the CGM is proportional to the mass in hot ($T>10^{5}$ K) supernova-driven outflows or the mass needed to balance gas consumption by star formation and gas ejection by the resulting supernova feedback.
    These scaling relations are shown in the right panels of Figures \ref{fig:mstar-novi-r200c} and \ref{fig:mstar-movi}.
    \item Among the models and simulations we consider, the \emph{EAGLE} zoom simulations of \citet{Oppenheimer:2018vn} and the model of \citet{Qu:2018va} are in closest agreement with our measurements across the entire mass range. Both predictions have the right mass dependence, but tend to be slightly low. The TNG-100 simulation \citep{Nelson:2018wy} quantitatively agrees with the measurements at $\Mstar \gtrsim 10^{10}$ \Msun, but is substantially lower than the measurements for smaller stellar masses. However, the \citet{Nelson:2018wy} predictions may not be directly comparable with our measurements because {of differences in how \OVI\ is associated with galaxies: \citet{Nelson:2018wy} use 3D spatial information to make their associations while we use 2D spatial and redshift information.}
    Comparisons of these and other models with the measurements are shown in Figures \ref{fig:mstar-novi-150}, \ref{fig:mstar-novi-r200c}, and \ref{fig:mstar-movi}.
\end{enumerate}

An important caveat about our measurements of the \OVI\ abundance is that they may be biased high by our assumption that the \OVI\ we detect is associated with the CGM of our chosen host galaxy.
It is possible that some of the \OVI\ appears in that particular galaxy's velocity window only coincidentally and is actually well outside the galaxy's CGM \citep[e.g.,][]{Ho:2021we}.

\subsubsection{The mass and column density of \OVI\ as a function of galaxy stellar mass}
\label{analysis:mass-dependence:abundance}
We summarize the \OVI\ content of a galaxy's CGM using three related quantities: the \OVI\ mass within a virial radius of the galaxy, $M_{\OVI}$; the area-weighted average column density within a virial radius, \ApNOVI{\RCCc}; and the area-weighted average column density within 150 kpc, \ApNOVI{\RCCc}.
For azimuthally symmetric \NOVI$(\IP)$, \ApNOVI{{\IPmax}} is given by the expression
\begin{equation}
    \label{eqn:apnovi}
    \ApNOVI{{\IPmax}}=
    \frac{2}{{\IPmax^2}}
    \int_{0}^{\IPmax} \NOVI(\IP') \IP' \dd \IP'.
\end{equation}
This expression does not include the contribution of the background \NOVI\ term in Equation \ref{eqn:expon-prof}.
$M_{\OVI}$ is defined to be $\ApNOVI{\RCCc} \times \pi \RCCc^2 \times m_{\rm O}$.

The summary quantities \ApNOVI{\rm 150 kpc}, \ApNOVI{\RCCc}, and $M_{\OVI}$ are shown in Figures \ref{fig:mstar-novi-150}, \ref{fig:mstar-novi-r200c}, and \ref{fig:mstar-movi}, respectively, and are listed in Table \ref{tab:mass-summaries}.
All three quantities reflect different combinations of the halo size and the \OVI\ volume density within the CGM (\nOVI).
$M_{\OVI}$ is an estimate of the total \OVI\ content of a halo.
Because this $M_{\OVI}$ is derived from column density measurements that are integrated along the line of sight, it may include some \OVI\ that is outside \RCCc\ in three dimensions but has an \IP\ that is within \RCCc.
\ApNOVI{\RCCc} is proportional to $M_{\OVI}/\RCCc^{2}$, and as a result depends less strongly on halo size than $M_{\OVI}$ does.
\ApNOVI{{\rm 150 kpc}} weights different parts of the halo differently for galaxies of different masses.
It is an average over the inner part of a $10^{10.5}$ \Msun\ galaxy halo, but an average over the entirety of a $10^{8.5}$ \Msun\ halo.

In addition to the measurements, Figures \ref{fig:mstar-novi-150}, \ref{fig:mstar-novi-r200c}, and \ref{fig:mstar-movi} show predictions and scaling relations for \OVI\ summary quantities as a function of galaxy mass.
\autoref{fig:mstar-novi-r200c} and \autoref{fig:mstar-movi} also include $M_{\OVI}$ and \ApNOVI{\RCCc}\ tracks corresponding to spheres with fixed \nOVI\ and radius \RCCc.
The fixed-\nOVI\ tracks are shown as gray lines in the right panels of these figures.
Some of the observed \OVI\ at $\IP<\RCCc$\ is likely to be outside \RCCc\ in three dimensions.
As a result, the average \nOVI\ within \RCCc\ of the observed galaxies will be lower than that of a fixed \nOVI\ track with similar $M_{\OVI}$ or \ApNOVI{\RCCc}.
If the fraction of the projected \NOVI\ that comes from outside \RCCc\ is relatively constant with host galaxy mass, then the fixed-\nOVI\ tracks can still serve as a guide for comparing the relative average \nOVI\ of galaxies with different masses.

\begin{deluxetable*}{lccc}
 \tabletypesize{\footnotesize}
 \tablewidth{\linewidth}
\tablecaption{
CGM \OVI\ content of star-forming galaxies
\label{tab:mass-summaries}}
\tablecolumns{4}
\tablehead{\colhead{$\logten$ Galaxy mass range} & \colhead{$\logten$ \ApNOVI{{\rm 150 kpc}}} & \colhead{$\log_{10}$ \ApNOVI{\RCCc}} & \colhead{$\log_{10}$ $M_{\OVI}$} \\
\colhead{[\Msun]} & \colhead{[\cmmt]} & \colhead{[\cmmt]} & \colhead{[\Msun]}}
\startdata
$7.8 \rng 8.5$ & $13.4^{+0.2 }_{-0.3 }$ & $13.9^{+0.2 }_{-0.3 }$ & $5.2^{+0.2 }_{-0.2 }$ \\[3pt]
$8.5 \rng 9.2$ & $13.7^{+0.2 }_{-0.4 }$ & $14.0^{+0.2 }_{-0.4 }$ & $5.6^{+0.1 }_{-0.4 }$ \\[3pt]
$9.2 \rng 9.6$ & $14.2^{+0.2 }_{-0.1 }$ & $14.4^{+0.1 }_{-0.1 }$ & $6.1^{+0.1 }_{-0.1 }$ \\[3pt]
$9.6 \rng 10.0$ & $14.3^{+0.2 }_{-0.1 }$ & $14.4^{+0.1 }_{-0.1 }$ & $6.3^{+0.1 }_{-0.2 }$ \\[3pt]
$10.0 \rng 10.4$ & $14.4^{+0.0 }_{-0.2 }$ & $14.3^{+0.1 }_{-0.1 }$ & $6.3^{+0.1 }_{-0.1 }$ \\[3pt]
$10.4 \rng 11.2$ & $14.3^{+0.1 }_{-0.3 }$ & $14.1^{+0.1 }_{-0.3 }$ & $6.5^{+0.1 }_{-0.2 }$ \\[3pt]
\enddata
\tablecomments{Summaries of the measured CGM-wide \OVI\ content of SF galaxies in different mass ranges.
The quantities are: the logarithm of the average \NOVI\ at $\IP<150$ kpc ($\logten$ \ApNOVI{{\rm 150 kpc}}); the logarithm of the average \NOVI\ at $\IP<\RCCc$ ({$\log_{10}$ \ApNOVI{\RCCc}}); and the logarithm of the total \OVI\ mass within \RCCc ({$\log_{10}$ $M_{\OVI}$}).
Values are given as a point estimate (median of the distribution) with asymmetric uncertainties (16th and 84th percentiles).
Uncertainties are calculated by propagating uncertainties on parameters of our model for $\NOVI(\Mstar,\IP)$ (see \S \ref{sec:analysis:sf-gal-model}).
}
\end{deluxetable*}

Figure \ref{fig:mstar-novi-r200c} shows our first conclusion for this section: the average \NOVI\ within a virial radius, \ApNOVI{\RCCc}, peaks at $\Mstar=10^{9.2}$ to 10$^{10}$ \Mstar.
The average equivalent \nOVI\ peaks over the same mass range.
Figure \ref{fig:mstar-movi} shows that the \OVI\ mass is monotonically increasing over the entire mass range we consider.
This monotonic increase is in part the result of more massive halos simply being larger.

\subsubsection{Scaling relations for $M_{\OVI}$}
\label{sec:analysis:mass-dependence:scalings}
We have found three scaling relations that adequately describe the \Mstar\ dependence of the (equivalent) summary quantities \ApNOVI{\RCCc}\ and $M_{\OVI}$.
These scaling relations were derived by examining different simple combinations of potentially relevant variables.
The relations are meant to be suggestive of {possible scenarios for \OVI\ generation,} but do not include a proper treatment of any relevant physical processes.
{Possible consequences of a more complete treatment of the physical processes are discussed in \ref{sec:discussion:implications}.}
{
The scenarios are: a collisionally ionized, ambient CGM phase at roughly the halo virial temperature; cooling hot outflows; and inflows that balance the rate at which gas is consumed and ejected by feedback.
}

The scaling relations are shown in the right panels of Figures \ref{fig:mstar-novi-r200c} and \ref{fig:mstar-movi}.
We give expressions for $M_{\OVI}$ as a function of $\Mstar$ below.
The corresponding $\ApNOVI{\RCCc}$\ can be found by dividing $M_{\OVI}$\ by $m_{\rm O} \pi \RCCc^2$, where $m_{\rm O}$ is the mass of an oxygen atom.
The scaling relations are:
\begin{align}
     M_{\OVI} &= \MCCc \left(\Omega_{b}/\Omega_{0}\right) f_{\rm CGM}\, f_{\rm O} \, \langle f_{\OVI} \rangle_{\TCCc(\MCCc)}
    \label{eqn:ambient-scaling}\\
    M_{\OVI} &= 10^{5.7}\, \Msun \, \left( \eta_{M, {\rm hot}} \times \frac{\text{SFR}}{\Msun \, \text{yr}^{-1}}\right)
    \label{eqn:outflow-scaling}\\
    M_{\OVI} &= 10^{5.3}\, \Msun \, \left(\left(1+\eta_{M, {\rm total}}\right) \times \frac{\text{SFR}}{\Msun \, \text{yr}^{-1}}\right)
    \label{eqn:inflow-scaling}
\end{align}
In the legends to Figures \ref{fig:mstar-novi-r200c} and \ref{fig:mstar-movi}, the relations are referred to as "Ambient", "Outflow," and "Balance," respectively.

In the ambient scaling relation, $\Omega_{b}/\Omega_{0}$ is the cosmic baryon fraction ($\approx 15$ \%), $f_{\rm CGM}$ is the fraction of potentially available baryons found in an ambient CGM phase, and $f_O$ is the mass fraction of oxygen among ambient CGM baryons.
We take $f_O$=0.001647, corresponding to a metallicity of 0.3 times solar given the solar abundances assumed  in \citet{Oppenheimer:2013vs}.
$\langle f_{\OVI} \rangle$ is the average fraction of oxygen in the form of \OVI, where the average is taken over a (base ten) log-normal temperature distribution centered on the halo virial temperature:
\begin{equation}
    \langle f_{\OVI} \rangle = \int_{-\infty}^{+\infty} f_{\OVI}(\logten T) p(\logten T)\, \dd \logten T.
\end{equation}
We assume the gas is in collisional ionization equilibrium and adopt the 0.3 times solar metallicity ion fraction tables from \citet{Oppenheimer:2013vs}.
The temperature distribution, $p(\logten T)$, is a Gaussian centered on $\logten \TCCc$\ with standard deviation $\sigma_{\logten T}$.
Taking the assumed metallicity of 0.3 times solar to be a reasonable fiducial value, the ambient scaling relation has two free parameters: $f_{\rm CGM}$ and $\sigma_{\logten T}$.
We use the values $f_{\rm CGM}=35$\% and $\sigma_{\logten T}=0.3$.
Both $f_{\rm CGM}$ and $\sigma_{\logten T}$ can be varied about these values while still producing $M_{\OVI}(\Mstar)$ relations that agree with the measurements.

In the outflow and balance scaling relations, the $\eta$ values are stellar-mass dependent mass loading factors for supernova-driven outflows as derived from the \emph{FIRE2} simulations by \citet{Pandya:2021vk}.
$\eta_{M, \rm{hot}}$ is the mass loading factor for gas with $T>10^{5}$ K; $\eta_{M, \rm{total}}$ is the mass loading factor for gas of all temperatures.
These mass loading factors give the amount of outflowing gas at $0.1 \rng 0.2$ $\rm R_{vir}$ that has enough kinetic energy to reach at least 0.5 $\rm R_{vir}$.
While the mass loading factors are calibrated using complex galaxy simulations, \citet{Pandya:2021vk} provide simple expressions for them in terms of galaxy \Mstar: $\eta_{M, \rm{total}}=10^{4.3}\left(\Mstar/\Msun\right)^{-0.45}$ and $\eta_{M, {\rm hot}}=10^{2.4}\left(\Mstar/\Msun\right)^{-0.27}$.
{The expression for $\eta_{M, {\rm total}}$ is given in their Equation 15.
The expression for $\eta_{M, {\rm hot}}$ is the product of Equation 15 and an expression for the fraction of the total ejected mass that is at $T>10^{5}$ K given in Equation 18.
}

The outflow and balance scaling relations include a dependence on the host galaxy star formation rate (SFR).
Our photometric SFR measurements are sufficient to allow a classification of galaxies as star-forming or quiescent (see end of \S \ref{sec:data:galaxies:cgmsq}), but not precise and accurate enough to use individually.
Instead, we assume the SF galaxy sample lies on the star-forming main sequence (SFMS) and use that to make a rough SFR estimate.
We use the lowest-redshift SFMS from \citet{Whitaker:2014tq}.
Using a different SFMS will change the shape and level of the scalings by the same factor.

\subsubsection{Predictions for \OVI\ from models and simulations}
\label{analysis:mass-dependence:models}
How well do predictions for the CGM \OVI\ content as a function of host galaxy stellar mass agree with our measurements?
We consider predictions from two categories of sources: simple single-halo models and cosmological hydrodynamic simulations\footnote{This does leave out single-halo simulations, which are known to have qualitative differences from cosmological simulations \citep{Fielding:2020vf}. However, we have not been able to find single-halo simulation publications that make predictions for a range of stellar masses.}.
We look at predictions from two models and three simulations and find that the best matches come from the \citet{Qu:2018va} model and the \citet{Oppenheimer:2018vn} \emph{EAGLE} zoom simulations.

The single-halo models are built by assuming a structure for the temperature, density, and metallicity of the CGM as a function of host galaxy mass and radius, assuming a dominant ionization mechanism, and calculating the resulting ionization structure.
We consider two sets of models: those of \citet{Qu:2018va} and those of \citet{Voit:2019tv}.
In the \citet{Qu:2018va} model, the CGM is dynamic, constantly cooling to balance gas consumption by star formation.
The gas is ionized by collisional and photoionization processes.
In the \citet{Voit:2019tv} model, the CGM is in hydrostatic equilibrium and the gas cooling time is 10 times the free fall time; the halo structure is static and persistent.
We are using the baseline model of \citet{Voit:2019tv} and do not include the log-normal temperature distribution about this baseline that is also explored in that work.
The gas is ionized by collisional processes only.
The authors of \citet{Qu:2018va}
Z. Qu (private communication) provided us with \ApNOVI{\RCCc} values calculated from their model.
\citet{Voit:2019tv} tabulate CGM structure parameters (Table 1 of their Appendix), from which we calculate our three \OVI\ content summaries.

We examine predictions from three sets of simulations: TNG-100 \citep{Nelson:2018wy}, \emph{EAGLE} zoom simulations \citep{Oppenheimer:2018vn}, and the full \emph{EAGLE} volumes \citep{Wijers:2020vr}.
\citet{Nelson:2018wy} and \citet{Wijers:2020vr} are self-consistent cosmological volumes; \citet{Oppenheimer:2018vn} is a set of re-simulations of particular \emph{EAGLE} halos with higher resolution and a non-equilibrium ionization module.
The TNG-100 and \emph{EAGLE} simulations are tuned to reproduce many of the same galaxy population observables, but have different implementations of astrophysical processes such as feedback.
\citet{Nelson:2018wy} provide tables of \ApNOVI{\RCCc} and $M_{\OVI}$.
Their masses are defined to only include gas that is gravitationally bound to the halo it is assigned to and as a result may not be directly comparable to our measurements.
\ApNOVI{\RCCc} is defined to be proportional to $M_{\OVI}/(\pi \RCCc^2)$ and so inherits the same selection effect.
The two \emph{EAGLE} simulation analyses, \citet{Oppenheimer:2018vn} and \citet{Wijers:2020vr}, calculate \NOVI\ in cylindrical slabs of fixed thickness around a galaxy without doing a three-dimensional pre-association and so are more directly comparable with observations; \citet{Oppenheimer:2018vn} use a slab thickness of 2400 kpc, \citet{Wijers:2020vr} use a slab thickness of 6.25 Mpc but then restrict to velocity windows around a galaxy of $\pm 300$ km s$^{-1}$.
\citet{Oppenheimer:2018vn} calculate \ApNOVI{\text{150 kpc}} for individual galaxies; B. Oppenheimer (private communication) provided us with these values.
\citet{Wijers:2020vr} show \NOVI\ as a function of radius and galaxy stellar mass (see their Figure 10).
We use these values to calculate all three \OVI\ summaries.

The measurements and predictions are shown in Figures \ref{fig:mstar-novi-150}, \ref{fig:mstar-novi-r200c}, and \ref{fig:mstar-movi}.
None of the predictions are in perfect agreement with the measurements.
However, two of the predictions roughly agree while three substantially disagree.
The three predictions that substantially disagree are: our implementation of the \citet{Voit:2019tv} model, the TNG-100 simulation values from \citet{Nelson:2018wy}, and the full-volume \emph{EAGLE} simulation values we calculated based on \citet{Wijers:2020vr}.
{The TNG-100 values agree with our measurements for the 2-3 highest mass bins, but disagree with the low-mass galaxy measurements.
Our implementation of \citet{Voit:2019tv} underpredicts all of the measurements.}
The full-volume \emph{EAGLE} simulations substantially underpredict the measurements at all stellar masses; they also underpredict the \emph{EAGLE} zoom simulations.
This difference has been ascribed to differences in resolution \citep{Oppenheimer:2016wy}.
\citet{Wijers:2020vr} note that \OVI\ column densities above $10^{14}$ \cmmt\ may not be fully converged at the resolution of the full volume simulations.
{All three of these sets of predictions differ from the measurements in shape as well as normalization: they overpredict the stellar mass of peak \OVI.}

The predictions that roughly agree are the \citet{Qu:2018va} model and the \emph{EAGLE} zoom simulation from \citet{Oppenheimer:2018vn}.
The \citet{Qu:2018va} values are slightly too low at $\Mstar>10^{9}$ \Msun, and the \emph{EAGLE} zoom simulation values are consistently a few tenths of a dex too low.
However, both predictions have approximately the right mass dependence.
In particular, and unlike the \citet{Voit:2019tv} and TNG-100 values, they do not have a steep drop in \OVI\ abundance below the stellar mass of peak \OVI.

\section{Discussion}
\label{sec:discussion}

\subsection{Robustness of the low-mass galaxy \NOVI\ estimates}
\label{sec:discussion:low-mass-association}

A key assumption of our analysis is that the \OVI\ absorbers we assign to galaxies arise in gas that is associated with those particular galaxies.
We make this assumption even though there is known to be an enhancement of \OVI\ incidence around galaxies due to galaxy clustering \citep{Finn:2016tq,Prochaska:2019vy}.
Put another way, we assume that we are measuring a 1-halo \OVI~term (due to the CGM of the assigned host galaxy) and possibly neglecting the contribution of a 2-halo \OVI~term (due to galaxy clustering).

The degree to which 2-halo term contamination will bias our 1-halo term estimate depends on the relative amplitudes of the terms.
\citet{Prochaska:2019vy} estimate that the covering fraction of 2-halo $\NOVI \geq 10^{13.5}$ \cmmt\ absorbers within 150 kpc of a galaxy should be about 10-20\%.
Examining \autoref{fig:expfit}, an extra $10^{13.5}$ \cmmt\ would constitute $1/10$th of a strong \OVI\ absorber associated with a $\Mstar \geq 10^{9.2}$ \Msun\ galaxy, but could be $1/3$rd or more of a strong lower-mass galaxy \OVI\ absorber.
This raises the question of whether we have (perhaps substantially!) overestimated the \OVI\ content of sub-$10^{9.2}$ \Msun\ galaxies.

The 2-halo term is statistical.
It is a convolution of 1-halo contributions arising from individual galaxies with a probability distribution over the separations between galaxies.
A specific case, such as a low-mass galaxy to which we have associated some column density of \OVI, is a realization from this distribution: either there is a contaminating galaxy present or there is not.
We believe we can rule out a scenario in which all or most of the \OVI\ we associate with low-mass galaxies actually arises in higher-mass galaxies.

These contaminating galaxies could be detected or they could have been missed by their source galaxy surveys.
Our main literature source of low-mass galaxies, \citet{Johnson:2017tp}, specifically selected field galaxies with no $L>0.1\, L^*$ neighbors.
39 of the 44 \CGMsq\ low-mass galaxies have no known $\Mstar \geq 10^{9.5}$ \Msun\ neighbors within $\pm 600$ km s$^{-1}$.
The 5 that do have a massive neighbor are all more than \RCCc\ from the sightline; only one of the five is a detection.
We can conclude that if there is 2-halo contamination, it is not coming from detected massive galaxies.

Whether contaminating galaxies with $\Mstar\geq 10^{9.2}$ \Msun\ have been missed will depend on the spectroscopic redshift survey completeness.
37 of the 44 \CGMsq\ low-mass galaxies are found in fields in which spectra were taken of \emph{all} targets within 2 arcminutes of the sightline and with $i$-band magnitude less than 22 \citepalias{Wilde:2021vr}.
It is therefore likely that the overwhelming majority of $\Mstar \geq 10^{9.2}$ \Msun\ galaxies that lie within the redshift range of the low-mass galaxies, $z=0.14\rng0.52$, and within 400 kpc of the sightline are, in fact, detected.
At least in the \CGMsq\ part of the low-mass galaxy sample, there should also be little to no 2-halo contamination from undetected, massive galaxies.

A final argument against low-mass galaxy \OVI\ being incorrectly ascribed high-mass galaxy \OVI\ is that in that situation, there should actually be more \OVI\ than we measure around the low-mass galaxies.
The typical \NOVI\ column density within \RCCc\ of a $\Mstar \geq 10^{9.2}$ \Msun\ galaxy is greater than the strongest \NOVI\ detection in our lowest-mass galaxy bin, $\Mstar=10^{7.8}\rng10^{8.5}$ \Msun.
If undetected high-mass galaxies are present near the galaxies in this bin, then either they must have atypically low \NOVI\ or they must somehow avoid being within \RCCc\ of the sightline.

These arguments do not rule out all possible versions of 2-halo term contamination of our estimate of the \OVI\ content of low-mass galaxies.
There may, for example, be an \OVI\ contribution from other, undetected, low-mass galaxies or the enriched IGM.
Even in that case, however, the \OVI\ would be coming from the CGM of a low-mass galaxy---the qualitative statement that low-mass galaxies have a detectable \OVI-traced CGM component is robust.

\subsection{The contribution of star-forming galaxy \OVI\ to \dNdz}
\label{sec:discussion:dNdz}

\begin{figure}
  \includegraphics[width=\linewidth]{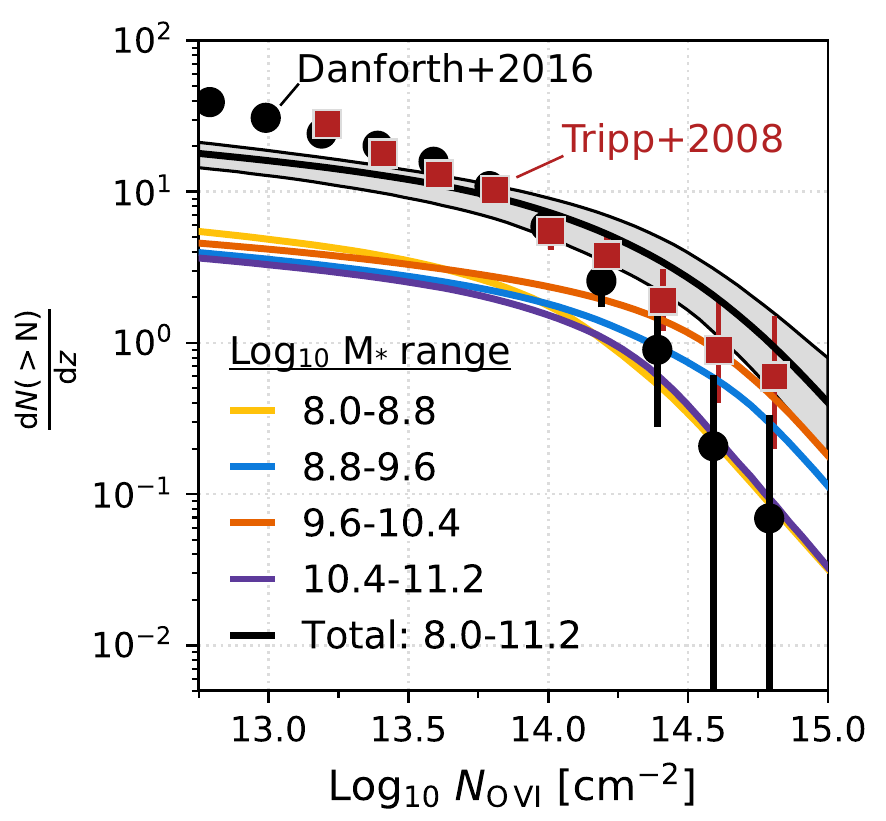}
  \caption{
  Cumulative distribution of \OVI\ system column densities as measured by galaxy-agnostic surveys and as predicted from the convolution of N$_{\OVI}(\Mstar,\IP)$ with a star-forming galaxy stellar mass function \citep{Tomczak:2014wi}.
  Curves show the contribution of star-forming galaxy gaseous halos to the \OVI\ incidence rate.
  Colorful curves split this contribution up in bins of stellar mass.
  The black curve is the median contribution of all the mass bins.
  The gray region represents the $1\sigma$-equivalent uncertainty in our empirical model for the distribution of \OVI\ around galaxies but does not include uncertainties in the stellar mass function.
  Points and squares are observational \dNdz\ measurements from \citet{Tripp:2008tj} and \citet{Danforth:2016uj}.
  Star-forming galaxies with \Mstar\ between $10^{8}$ and $10^{11.2}$ \Msun\ can explain most \OVI\ systems with column densities greater than 10$^{13.5}$ \cmmt.
  }
  \label{fig:dNdz}
\end{figure}

This study measures how \NOVI\ around individual star-forming galaxies depends on galaxy mass and impact parameter.
Other studies have measured the incidence of \OVI\ along random sightlines, \dNdz\  \citep{Tripp:2008tj,Danforth:2016uj}.
We can use our individual-galaxy measurements to estimate the contribution of the galaxies we are studying to the general \OVI\ incidence.
Comparing the measured general \OVI\ incidence with our estimate provides a check on our individual-galaxy measurements---there cannot be more \OVI\ around a subset of the galaxy population than there is around all galaxies.
The three results of this section are:
\begin{enumerate}
    \item SF galaxies with masses between $10^{8}$ and $10^{11.2}$ \Msun\ can account for most absorption systems with \NOVI\ greater than $10^{13.5}$ \cmmt.
    \item SF galaxies with masses between $10^{8.8}$ and $10^{10.4}$ \Msun\ can account for all absorption systems with \NOVI\ greater than $10^{14.5}$ \cmmt.
    \item An additional source of \OVI\ may be necessary to account for absorption systems with $\NOVI<10^{13.25}$ \cmmt.
    \item The modal $\NOVI>10^{13.5}$ \cmmt\ absorber is associated with a \Mstar$\approx 10^{10}$ \Msun\ galaxy.
\end{enumerate}

To estimate the \OVI\ incidence due to galaxies in our study, we combine a stellar mass function with the parametric description of \NOVI\ as a function of stellar mass and impact parameter from Section \ref{sec:analysis}.
Through this combination, we calculate the expected differential number of absorption systems with \OVI\ column densities greater than some threshold \NOVI, $\frac{\dd N}{\dd z}(z, \NOVI)$.
Our expression for the combination of the stellar mass function and \NOVI\ distribution is a modified version of the absorption distance \citep{Bahcall:1969tt,Hogg:1999tj}:
\begin{equation}
    \label{eqn:dNdz}
    \begin{split}
        \frac{\dd N}{\dd z}(z,& \NOVI) = 2\pi (1+z)^2 \frac{D_H}{E(z)}\\
        &\times \int_{\mathcal{M}_{min}}^{\mathcal{M}_{max}}
            \int_{0}^{R_{\perp,\, max}} \Phi(z, \mathcal{M}) \\
        & \times p(N>\NOVI \vert \mathcal{M}, \IP) \IP \dd \IP \dd \mathcal{M}.
    \end{split}
\end{equation}
$\mathcal{M}$ is $\lten (\Mstar/\Msun)$, $D_H=c/H_0$ is the Hubble distance, and $E(z)=H(z)/H_0$ is the redshift scaling of the Hubble parameter.
$\Phi(z, \mathcal{M})$ is a stellar mass function; we use the star-forming galaxy mass function of \citet{Tomczak:2014wi}.
The factors of $(1+z)^2$ are included because we use proper (non-comoving) \IP\ in our calculations.
$p(N>\NOVI \vert \mathcal{M}, \IP)$\ is calculated by integrating \autoref{eqn:laplace-dist} from $\NOVI$ to infinity, meaning that this expression includes the contribution of intrinsic scatter at fixed $\mathcal{M}$ and \IP.
We evaluate this expression for multiple realizations of our \NOVI\ parametric model to propagate the uncertainty on that model's parameters through to \dNdz.

The \dNdz\ estimate and its uncertainty are shown in Figure \ref{fig:dNdz}.
For \NOVI\ below $10^{13.25}$ \cmmt, there is a clear need for an additional source of \OVI\ absorbers.
Higher column density \OVI\ systems are consistent with being almost entirely due to these SF galaxies.
The highest column density \OVI\ systems ($>10^{14.5}$ \cmmt) are consistent with being mostly due to SF galaxies with \Mstar\ between $10^{8.8}$ and $10^{10.4}$ \Msun---galaxies with twice the stellar mass of the Small Magellanic Cloud ($10^{8.5}$ \Msun, \citealt{Skibba:2012uj}) to galaxies with half the stellar mass of the Milky Way ($10^{10.7}$ \Msun, \citealt{Bland-Hawthorn:2016vl}).

At $\NOVI \geq 10^{14.25}$ \cmmt, our \dNdz\ estimates are greater than the \citet{Danforth:2016uj} measurement by more than the nominal uncertainties.
This disagreement is most likely due to differences in how we and \citet{Danforth:2016uj} combine absorption components to form absorption systems.
We define an absorption system to be the absorption components within $\pm 300$ km s$^{-1}$ of a galaxy.
\citet{Danforth:2016uj} define two components to be part of the same system if they are within 30 km s$^{-1}$ (or a larger velocity span for stronger, broader lines) of each other.
In their Figure 2, they show three \OVI\ absorption components with $z=0.12363, 0.12389,$ and 0.12479, corresponding to a total velocity span of 310 km s$^{-1}$.
By their definition, these components are part of three separate systems.
By our definition, they could be part of a single absorption system.
\citet{Danforth:2016uj} will tend to split what we would call a single high column density system into several lower column density systems, reducing their measured high column density \dNdz.

Where could the additional low column density \OVI\ absorbers come from?
One possibility is that this gas comes from the farther outskirts of galaxies in our sample.
Most of our survey is only complete to a column density of about $10^{13.8}$ cm$^{-2}$.
We would therefore not  be able to detect an extended $\NOVI<10^{13.25}$ \cmmt\ envelope around galaxies, if such an envelope were present.
Extended \OVI\ distributions of this sort are seen, for example, in simulations of dwarf galaxies \citep{Mina:2020tx}.
Other possible sources include dwarf galaxies with masses below $10^8$ \Msun, quiescent galaxies, galaxy groups and clusters, and the IGM \citep{Prochaska:2011vo,Finn:2016tq,Bielby:2019vv,Burchett:2018tn,Stocke:2019vr}.

\subsection{The distribution of \OVI\ around star-forming galaxies}
\label{sec:discussion:OVI-distribution}
Where, relative to the central galaxy, is the \OVI\ located?
Our dataset does not provide an insight into departures from isotropy, but is informative about the typical \OVI\ radial distribution.

Detectable \OVI\ absorption (e.g., \NOVI$\gtrsim 10^{14}$ \cmmt) extends to about \RCCc, as can be seen from our fits to \NOVI\ as a function of stellar mass and impact parameter (see \S \ref{sec:analysis:sf-gal-model} and Figure \ref{fig:expfit}).
The \OVI\ CGM of galaxies at the stellar mass of peak \OVI\ abundance ($\Mstar \sim 10^{9.5}$ \Msun) has a slightly greater extent, in units of \RCCc, than that of dwarf galaxies or $L^*$ or super-$L^*$ galaxies.
This extent is consistent with previous studies of the radial distribution of \OVI\ (e.g., \citealt{Prochaska:2011vo,Johnson:2015tj}).

An estimate of the volume density profiles $\nOVI(\Mstar, \IP)$ can be calculated from $\NOVI(\Mstar, \IP)$ using the inverse Abel transform.
This estimate explicitly assumes that $\nOVI$ is isotropic.
Integrating the volume density profiles out to different radii provides information on the \OVI\ mass distribution as a function of radius.
We find that about 60\% of the \OVI\ mass is found within \RCCc and 95\% is found within $2\RCCc$ of the galaxy.
This extent is approximately consistent with analyses of the \OVI\ distribution in simulations \citep{Oppenheimer:2016wy,Strawn:2021ua,Ho:2021we}.

This result is model dependent.
\OVI\ absorption outside \RCCc is mostly below the detection limit, meaning that the shape of $\NOVI(\IP)$ past \RCCc\ is only weakly constrained.
We have assumed that \NOVI\ is a declining exponential function of \IP.
This assumption provides an adequate description of the measurements down to $\NOVI \approx 10^{13.5}$ \cmmt.
It could be the case that $\NOVI(\IP)$\ changes to a shallower function of \IP, such as a power law, at $\NOVI$ values below our detection limit.
If this change were to occur, then a greater fraction of the \OVI\ mass would be found outside \RCCc.
{We have also assumed that \NOVI, and hence \nOVI, is isotropic.
As is quantified in \citet{Strawn:2021ua}, this assumption can lead to underestimation of how much mass is present outside of any given radius.}

Prior observational studies of the distribution of \OVI\ around galaxies from COS-Halos and eCGM found a half-mass radius of about 0.6 times the virial radius \citep{Mathews:2017uf,Stern:2018to}, while we find a half-mass radius of about 0.9 times the virial radius.
This difference may be the result of different modeling choices.
\citet{Stern:2018to} compute $\nOVI$ only as far as the virial radius, while we compute $\nOVI$ to greater distances.

Project AMIGA \citep{Lehner:2020tk}, a survey of a variety of ions including \OVI\ along 43 QSO sightlines around M31, provides a point of comparison for our findings.
\citet{Lehner:2020tk} find that \OVI\ absorption with $\NOVI > 10^{14}$ \cmmt\ is detected out to the highest-\IP\ sightline with spectral coverage of \OVI, $\IP \approx 560$ kpc (corresponding to $\IP/\RCCc = 2.5$).
The $\NOVI\geq 10^{14.6}$ \cmmt\ covering fraction at $\IP/\RCCc = 2.5$ is about $1/2$.
In our sample, star-forming galaxies with $\Mstar=10^{11}$ \Msun\ have median $\NOVI=10^{14.6}$ \cmmt\ at $\IP/\RCCc \lesssim 0.3$.
M31's \OVI-traced CGM is therefore significantly more extended than that of the majority of the galaxies in our sample.
Determining whether this greater extent is the result of M31's environment or a reflection of simple halo-to-halo variance in CGM properties will require multi-sightline studies of the CGM of other galaxies.

\subsection{The influence of galaxy environment}
\label{sec:discussion:environment}

\begin{figure}
    \centering
    \includegraphics[width=\linewidth]{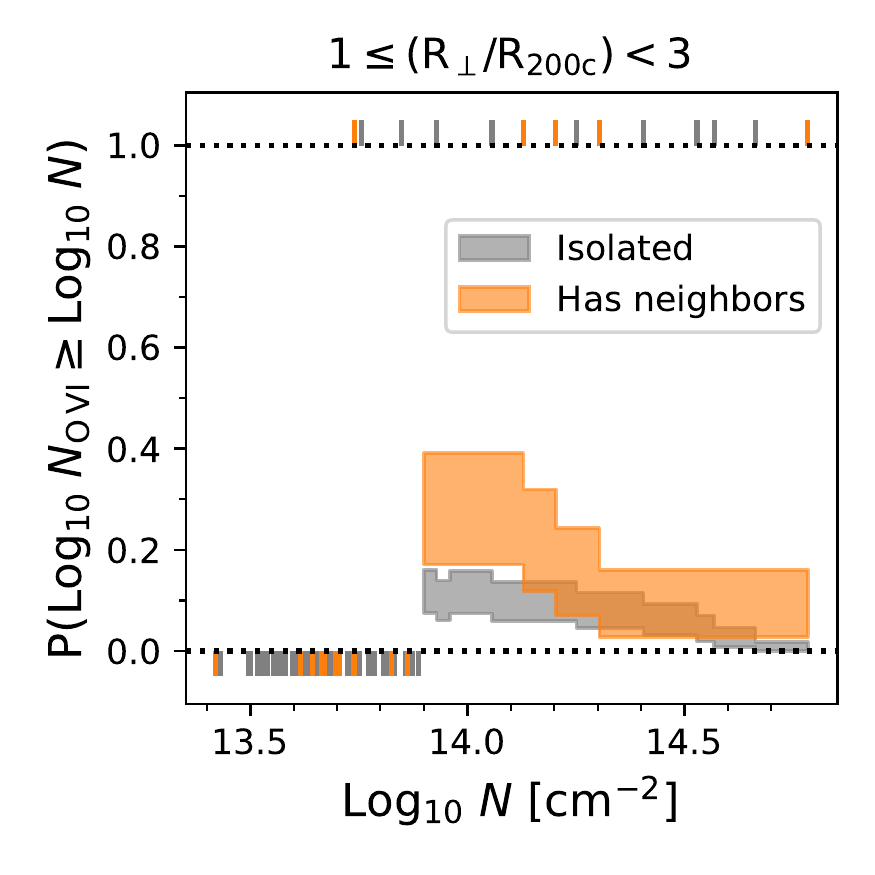}
    \caption{
    \OVI{} covering fractions of \CGMsq{} galaxies with and without neighbors with stellar mass greater than 10$^{9.5}$ M$_{\rm Sun}$.
    The figure shows covering fraction of the two sub-samples over a range of column density thresholds; the shaded regions are 68\% binomial credible intervals.
    The detected column densities and $3\sigma$ upper limits on non-detections are shown as ticks at the top and bottom, respectively, of the plot.
    Galaxies with neighbors have marginally higher \OVI{} covering fractions than galaxies without neighbors at column densities of about 10$^{14}$ cm$^{-2}$.
    }
    \label{fig:discussion:env-covering-fracs}
\end{figure}

An effect that has been noted in the literature is that galaxies with \emph{neighbors} have detectable absorption in ions including \specnotation{Mg}{II} and \OVI\ out to greater distances than galaxies without neighbors \citep{Johnson:2015tj,Dutta:2021vq}{, though this effect may not apply at all halo mass scales \citep{Pointon:2017wl}.}
We see this more extended \OVI\ around galaxies with neighbors in \CGMsq\ as well.
This effect may be responsible for some of the high-\OVI\ absorbers at impact parameters greater than \RCCc\ in \autoref{fig:expfit}.

The effect can be seen by comparing the \NOVI\ covering fractions of galaxy sub-samples with and without neighbors.
We restrict ourselves to the \CGMsq\ galaxies in our sample{; we do not include galaxies from any other sources in this analysis}.
We define a \emph{neighbor} to be a galaxy (star-forming or quiescent) with stellar mass greater than 10$^{9.5}$ \Msun\ that is within 400 kpc of the sightline and within 600 km s$^{-1}$ of the absorber host galaxy.
The mass requirement ensures that neighbors are detectable over the entire redshift range.
A host galaxy without any neighbors is isolated; a host galaxy with one or more neighbors is not isolated.

The difference between isolated and non-isolated galaxies is clearest for host galaxies that are 1 to 3 times \RCCc\ from the sightline.
Covering fractions for the isolated and non-isolated sub-samples at a range of \NOVI\ threshold values are shown in Figure \ref{fig:discussion:env-covering-fracs}.
The ranges shown are 68\% binomial confidence intervals calculated assuming a Jeffrey's prior on the covering fraction.
At limiting column densities less than $10^{14.1}$ cm$^{-2}$, the covering fraction for non-isolated galaxies is 1.5 to 5 times greater than that for isolated galaxies.
At greater limiting column densities, isolated and non-isolated galaxies have consistent covering fractions.
Isolated and non-isolated galaxies that are more than 3 times \RCCc\ from the sightline (not shown in \autoref{fig:discussion:env-covering-fracs}) have consistent covering fractions at all \NOVI\ thresholds.

{
These elevated covering fractions are consistent with the results of \citet{Johnson:2015tj}, but may or may not be consistent with the results of \citet{Pointon:2017wl}.
\citet{Pointon:2017wl} compare \OVI\ equivalent widths and covering fractions between isolated galaxies and galaxies in groups.
They find that the average \OVI\ equivalent width within 350 kpc is lower for group galaxies relative to isolated ones.
The covering fractions at an equivalent width of 0.06\AA\ are consistent within uncertainties, but the point estimate is lower for group galaxies.}

{
The mapping between the \citet{Pointon:2017wl} results and ours is not obvious.
Most of the \citet{Pointon:2017wl} detections are within 150 kpc while our analysis focuses on the region immediately outside a galaxy's \RCCc,  corresponding to a minimum physical $\IP$\ of $\approx 150$ kpc.
Even at $\IP>150$ kpc, isolated galaxies in \citet{Pointon:2017wl} have greater equivalent widths than group galaxies (see their Figure 3).
The \citet{Pointon:2017wl} analysis focuses on group mass halos while our neighbor-having galaxies are generally sub-group scale.
This difference in mass may be responsible for the difference in behavior of \OVI: perhaps having neighbors increases the \OVI\ incidence rate at lower masses and then decreases it once a neighborhood approaches the mass scale of a galaxy group.
}

Could the inclusion of neighbor-having galaxies in the general sample be biasing the \NOVI\ model of \S \ref{sec:analysis:sf-gal-model} to higher values or greater extents?
The model is designed to be robust to outliers.
In particular, we use a heavy-tailed distribution to describe the amount of scatter in \NOVI\ at fixed \Mstar\ and \IP.
As a test, we re-calculated the parameters of the column density profile model while excluding the non-isolated galaxies.
The resulting parameters are all consistent with the full-sample values given in \autoref{tab:exponential-model-param-results}.

The depth, completeness, and size of \CGMsq\ mean that it is well suited for investigating how \OVI\ incidence depends on more detailed metrics of sub-Mpc environment.
This investigation is beyond the scope of the present paper and is left for future work.

\subsection{What conditions does \OVI\ trace?}
\label{sec:discussion:implications}

In \S \ref{sec:intro}, we introduced three possible scenarios for \OVI\ generation: ambient halo gas, outflows, and inflows.
In \S \ref{sec:analysis:mass-dependence:scalings}, we defined three expressions for the mass of \OVI\ associated with a galaxy as a function of $\Mstar$ that can be interpreted in terms of these possible \OVI\ generation scenarios.
All three expressions are consistent with the measurements (see \autoref{fig:mstar-novi-r200c} and \ref{fig:mstar-movi}), suggesting that all three possible scenarios of \OVI\ generation are viable.
However, if the scaling relations are refined to be more physical, they become less viable as individually complete explanations of the observed \OVI.
In brief, the three scenarios have the following problems:
\begin{itemize}
    \item An ambient collisionally ionized phase: We have assumed that the CGM has a temperature distribution centered on the virial temperature. More physical models of ambient halo gas suggest that the temperature distribution of an ambient CGM phase would be preferentially below, rather than centered on, the virial temperature.
    \item Hot outflows: Half of the \OVI\ mass is outside 0.9\RCCc. By the time they reached these radii, hot outflows would cool to temperatures too low to produce \OVI\ through collisional ionization.
    \item Inflows: The required mass of \OVI\ per unit inflow mass would require near-ideal conditions for \OVI\ production.
\end{itemize}
Explaining the observed amount, extent, and host galaxy mass dependence of \OVI\ requires: (1) a different, more favorable physical refinement of one or more of the scenarios, (2) an evolution from one scenario to another with mass, or (3) a superposition of the scenarios at fixed mass.

In our expression for an ambient CGM phase, we assume that the gas temperature is distributed log-normally around \TCCc.
The fraction of oxygen in the form of \OVI\ ($f_{\OVI}$) in collisional ionization equilibrium peaks at $T_{\OVI}\equiv 10^{5.5}$ K.
The stellar mass where \TCCc\ is equal to $T_{\OVI}$ is $10^{9.6}$ \Msun, which is coincident with the peak of \ApNOVI{\RCCc}.
At lower and higher masses, the upper and lower tails of the log-normal distribution, respectively, are the source of the \OVI\ mass in the ambient scaling relation.

In more physically motivated models of the ambient CGM, temperatures are preferentially below \TCCc.
This is why the mass of peak \ApNOVI{\RCCc}\ in the \citet{Voit:2019tv} model is at greater masses than is observed: the mass of $T_{\OVI}$ gas is greater around galaxies with $\TCCc>T_{\OVI}$ than around galaxies with $\TCCc\approx T_{\OVI}$.
An analysis of CGM temperatures in high resolution hydrodynamic simulations by \citet{Lochhaas:2021to} finds a similar result: even in virialized galaxy-scale halos, gas on the outskirts of the CGM (i.e., where much of the mass resides) has a characteristic temperature of about $1/2\TCCc$, rather than $\TCCc$.
Halos in which some of the assembly energy is still in the form of coherent, non-thermal motions have an even lower characteristic temperature.
If the temperature distributions in these models are a more accurate representation of reality than our \TCCc-centered log-normal distributions, then a collisionally ionized ambient phase cannot explain much of the \OVI\ around galaxies with $\Mstar \lesssim 10^{10}$ \Msun.

Hot outflows cannot be the primary explanation for the observed \OVI\ because they would not be able to reproduce its extent.
An outflow starting out at temperatures greater than $T_{\OVI}$\ will eventually cool to temperatures too low to produce \OVI\ through collisional ionization, provided that it is not expanding into a medium with $T\geq T_{\OVI}$.
The details of how quickly this cooling will happen depend on the outflow model used.
In general, an outflow will cool to below $T_{\OVI}$\ after traveling 10-100 kpc from its starting point \citep{Thompson:2016vp,Keller:2020we}.
The extent of hot outflow \OVI\ would therefore be smaller than what is observed.

The argument against inflows as the dominant explanation for \OVI\ is the prohibitively large average fraction of oxygen in the form of \OVI\ that would be required.
Assuming a steady inflow $\dot{M}$ that takes $t_{flow}$ to arrive at the galaxy, the mass of \OVI\ associated with the inflow would be $\dot{M}\, t_{flow}\, f_{\rm O} \langle f_{\OVI}\rangle$, where $f_{\rm O}$ is the mass fraction of oxygen in the flow and $\langle f_{\OVI} \rangle$ is the fraction of oxygen in the form of \OVI\ averaged over the flow.
For a fiducial flow rate of 1 \Msun\ yr$^{-1}$, metallicity $0.3 Z_\odot$, and $t_{flow}=1$ Gyr and a maximal $\langle f_{\OVI} \rangle = 0.25$, the \OVI\ mass would be about $10^{5.6}$ \Msun.
The required amount is only $10^{5.3}$ \Msun, a factor of two lower than this fiducial value.

However, the value of $\langle f_{\OVI} \rangle$ used in this expression is maximal and applies over a narrow range in density and temperature.
Even in simulations in which \OVI\ is mostly found in photoionized inflows, $\langle f_{\OVI} \rangle$ is  at least 5 times smaller than this maximal value at large radii and declines steeply at smaller radii \citep{Strawn:2021ua}.
The metallicity may also be lower than $0.3 Z_\odot$, depending on the origin of the inflow: \citet{Strawn:2021ua} find average metallicities that are slightly smaller than $0.1 Z_\odot$.
Conversely, $t_{flow}$ could be greater than 1 Gyr.
The free-fall time from \RCCc, for example, is about 1.7 Gyr.
Taking an infall time of 3 Gyr, a reduction of $\langle f_{\OVI}\rangle$ by a factor of 5, and a reduction of $f_{\rm O}$ by a factor of 3, the \OVI\ mass associated with the flow drops to $10^{4.9}$ \Msun.
On balance, the \OVI\ mass associated with inflows is likely to be lower than what is needed by a factor of at least 2.

The \OVI\ observations can be explained by altering the CGM temperature structure away from hydrostatic equilibrium, by superimposing different scenarios, or by invoking an evolution of the dominant scenario with galaxy mass.
\citet{Voit:2019tv} and \citet{Roy:2021vv} extend their hydrostatic ambient model with a log-normal temperature distribution about the equilibrium temperature.
\citet{Qu:2018va} invoke radiative cooling from 2\TCCc.
Both approaches extend the range of galaxy masses over which the CGM can produce \OVI\ through collisional ionization.
\citet{Stern:2018to} instead assume that gas on the outskirts of the CGM has a low density and is in thermal equilibrium with the ultraviolet background, conditions that are favorable for the production of \OVI\ through photoionization.
\citet{Strawn:2021ua} find that \OVI\ is produced through different mechanisms in different parts of the CGM: an ambient, outflow-heated phase at small radii, photoionized inflows at larger radii.
\citet{Qu:2018va} and \citet{Roy:2021vv} find that there is a transition from cool, photoionized \OVI\ production at low \Mstar\ to collisional \OVI\ production in ambient gas at higher \Mstar.

Our measurements tentatively favor a superposition of an inner and an outer \OVI\ generation mechanism.
The large extent of the CGM suggests that the ``cool outer inflow" scenario of \citet{Stern:2018to} and \citet{Strawn:2021ua} is plausible.
The need to reproduce the inner half of the \OVI\ mass suggests a need for an ambient phase with some non-hydrostatic temperature structure.
Testing this tentative explanation will require combining measurements of \OVI\ with measurements of other highly ionized species, such as \specnotation{Ne}{VIII}.
A study of galaxy-\specnotation{Ne}{VIII}\ absorber pairs at $z=0.49\rng1.44$ found that the \specnotation{Ne}{VIII}\ was consistent with an origin in virial temperature ambient gas \citep{Burchett:2019vk}.
Studies of galaxy-\OVI\ absorber-\specnotation{Ne}{VIII}\ absorber triplets will provide a stronger constraint on the ionization mechanisms responsible for these species in different parts of the CGM.

\section{Conclusion}
\label{sec:conclusion}

We have collated a sample of 249 unique galaxy-\OVI\ absorber pairs, 126 of which were found as part of the \CGMsq\ survey and had not been published before.
The sample includes galaxies with stellar masses $\Mstar=10^{7.8}\rng10^{11.2}$ \Msun, redshifts $z=0.002\rng0.6$, and impact parameters $\IP=11\rng400$ kpc.

We used this sample to study how stellar mass and galaxy environment affect the incidence of \OVI\ absorption around star-forming galaxies.
We created a descriptive model for \NOVI(\Mstar, \IP) around star-forming galaxies and used this model to estimate the total \OVI\ content within a virial radius of a star-forming galaxy.
We compared these estimates with predictions from theoretical models and simulations and wrote down several scaling relations which provide an accurate description of a star-forming galaxy CGM's \OVI\ content as a function of \Mstar.
In addition to studying the \Mstar-dependence of \OVI\ absorption around star-forming galaxies, we also quantified the effects of the presence of $\Mstar \geq 10^{9.5}$ \Msun\ neighbors on \NOVI.

A methodological conclusion of this work is that analyses of CGM absorption need to \emph{simultaneously} control for the absorber host galaxy's \Mstar\ and \IP (and, possibly, other variables as well).
The strong dependence of \NOVI\ on both quantities means that comparisons that control for only one of the variables at a time risk confusing a trend in one of the quantities for a trend in the other.

We measured several quantitative properties of the distribution of \OVI\ around galaxies.
\begin{itemize}
    \item Star-forming $\Mstar=10^{9.2}\rng10^{10}$ \Msun\ galaxies have the greatest average \NOVI\ within \RCCc\ (\S \ref{sec:analysis:sf-mass-trend}, \autoref{fig:mstar-novi-r200c}).
    \item Galaxies with at least one $\Mstar \geq 10^{9.5}$ \Msun\ neighbor have a $\NOVI\geq 10^{14}$ \cmmt\ covering fraction that is 1.5 to 5 times greater than that of galaxies without such neighbors (\S \ref{sec:discussion:environment}).
    \item Detectable \OVI\ ($\NOVI \gtrsim 10^{14}$ \cmmt) extends to about $1\times\RCCc$ for star-forming galaxies with $\Mstar=10^{8.5}\rng10^{11.2}$ \Msun\ (\S \ref{sec:discussion:OVI-distribution}).
    Assuming an isotropic volume density distribution $\nOVI(\IP)$, 60\% of the total \OVI\ around a galaxy is found within a radius of \RCCc\ and 95\% is found within a radius of 2\RCCc.
\end{itemize}

Analyzing these measured quantities further, we derived a number of implications for the nature of \OVI\ and efforts to model it.
\begin{itemize}
    \item By combining our model for \NOVI(\Mstar, \IP) around star-forming galaxies with a stellar mass function, we calculated the expected contribution of star-forming galaxies in the mass range we consider to the cumulative column density distribution of \OVI\ systems, \dNdz. By comparing our calculation with \dNdz\ measurements from \citet{Danforth:2016uj} and \citet{Tripp:2008tj}, we found that our galaxies can account for the majority of \OVI\ systems with $\NOVI > 10^{13.5}$ \cmmt. Systems with $\NOVI < 10^{13.5}$ \cmmt\ require an additional source of \OVI\ absorption (\S \ref{sec:discussion:dNdz}).
    \item We {constructed} three simple scaling relations that accurately describe the mass of \OVI\ in a star-forming galaxy halo as a function of \Mstar. One relation assumes that \OVI\ is found in a collisionally ionized ambient phase with temperatures distributed around the virial temperature (\autoref{eqn:ambient-scaling}). The other two relations depend on the host galaxy star formation rate and on supernova outflow mass loading factors from \citet{Pandya:2021vk} (\autoref{eqn:outflow-scaling} and \autoref{eqn:inflow-scaling}).
    \item We compared predictions for different halo-scale summaries of \OVI\ content as a function of \Mstar\ with our measurements and found that the two predictions that come closest are the EAGLE zoom simulations of \citet{Oppenheimer:2018vn} and the model of \citet{Qu:2018va} (\S \ref{analysis:mass-dependence:models}).
    \item We argued that explaining the observed spatial extent and mass dependence of \OVI\ requires a combination of different \OVI\ generation mechanisms (\S \ref{sec:discussion:implications}) and/or departures from a hydrostatic equilibrium temperature structure for the CGM.
\end{itemize}

\begin{acknowledgements}
KT would like to thank Matt McQuinn and Yakov Faerman for useful discussions, Ben Oppenheimer and Zhijie Qu for providing predictions from their models, and the referee for a helpful and constructive report.
This research has made use of the HSLA database, developed and maintained at STScI, Baltimore, USA. KT, JKW, and MCW acknowledge support from from NSF-AST 1812521. Additionally, JKW and KT acknowledge support from the Research Corporation for Science Advancement, Cottrell Scholar grant ID number 26842. The ongoing efforts of the Student Quasar Absorption Diagnosticians (aka Werk SQuAD) in the University of Washington Astronomy Department to vet automated measurements of thousands of galaxy redshifts and vet Voigt profile fits for thousands of absorption lines in CGM$^{2}$ were instrumental to the science presented in this study.  This work benefited from lively Zoom discussions during KITP's ``Fundamentals of Gaseous Halos" program, and thus was supported in part by the National Science Foundation under Grant No. NSF PHY-1748958.

This work was based on observations obtained at the Gemini Observatory, which is operated by the Association of Universities for Research in Astronomy, Inc., under a cooperative agreement with the NSF on behalf of the Gemini partnership, whose members are listed at \url{https://noirlab.edu/science/about/scientific-acknowledgments#ack-gemini}. The Gemini North telescope is located within the Maunakea Science Reserve and adjacent to the summit of Maunakea. We are grateful for the privilege of observing the Universe from a place that is unique in both its astronomical quality and its cultural significance.

Funding for the Sloan Digital Sky Survey IV has been provided by the Alfred P. Sloan Foundation, the U.S. Department of Energy Office of Science, and the Participating Institutions. A listing of the participating institutions can be found at \url{https://www.sdss.org/collaboration/citing-sdss/}.

The Pan-STARRS1 Surveys (PS1) and the PS1 public science archive have been made possible through contributions by a variety of institutions, a listing of which can be found at \url{https://panstarrs.stsci.edu/}

The Legacy Surveys consist of three individual and complementary projects: the Dark Energy Camera Legacy Survey (DECaLS; NOAO Proposal ID  2014B-0404; PIs: David Schlegel and Arjun Dey), the Beijing-Arizona Sky Survey (BASS; NOAO Proposal ID  2015A-0801; PIs: Zhou Xu and Xiaohui Fan), and the Mayall z-band Legacy Survey (MzLS; NOAO Proposal ID 2016A-0453; PI: Arjun Dey). Details on the support of each constituent of the Legacy Surveys can be found at \url{https://www.legacysurvey.org/acknowledgment/}.
\end{acknowledgements}

\software{\texttt{astropy} \citep{Astropy-Collaboration:2013uv,Astropy-Collaboration:2018vm},
\\ \texttt{linetools} \citep{Prochaska:2017vh},
\\ \texttt{matplotlib}  \citep{Hunter:2007ux},
\\ \texttt{numpy} \citep{Harris:2020ti},
\\
\texttt{pandas} \citep{McKinney:2010vw},
\\
\texttt{pymc3} \citep{Salvatier:2016ui}
}

\bibliography{main.bbl}

\begin{thebibliography}{}
\expandafter\ifx\csname natexlab\endcsname\relax\def\natexlab#1{#1}\fi
\providecommand{\url}[1]{\href{#1}{#1}}
\providecommand{\dodoi}[1]{doi:~\href{http://doi.org/#1}{\nolinkurl{#1}}}
\providecommand{\doeprint}[1]{\href{http://ascl.net/#1}{\nolinkurl{http://ascl.net/#1}}}
\providecommand{\doarXiv}[1]{\href{https://arxiv.org/abs/#1}{\nolinkurl{https://arxiv.org/abs/#1}}}

\bibitem[{{Abolfathi} {et~al.}(2018){Abolfathi}, {Aguado}, {Aguilar}, {Allende
  Prieto}, {Almeida}, {Ananna}, {Anders}, {Anderson}, {Andrews}, {Anguiano},
  {Arag{\'o}n-Salamanca}, {Argudo-Fern{\'a}ndez}, {Armengaud}, {Ata},
  {Aubourg}, {Avila-Reese}, {Badenes}, {Bailey}, {Balland}, {Barger},
  {Barrera-Ballesteros}, {Bartosz}, {Bastien}, {Bates}, {Baumgarten},
  {Bautista}, {Beaton}, {Beers}, {Belfiore}, {Bender}, {Bernardi}, {Bershady},
  {Beutler}, {Bird}, {Bizyaev}, {Blanc}, {Blanton}, {Blomqvist}, {Bolton},
  {Boquien}, {Borissova}, {Bovy}, {Bradna Diaz}, {Brandt}, {Brinkmann},
  {Brownstein}, {Bundy}, {Burgasser}, {Burtin}, {Busca}, {Ca{\~n}as},
  {Cano-D{\'\i}az}, {Cappellari}, {Carrera}, {Casey}, {Cervantes Sodi}, {Chen},
  {Cherinka}, {Chiappini}, {Choi}, {Chojnowski}, {Chuang}, {Chung}, {Clerc},
  {Cohen}, {Comerford}, {Comparat}, {Correa do Nascimento}, {da Costa},
  {Cousinou}, {Covey}, {Crane}, {Cruz-Gonzalez}, {Cunha}, {da Silva Ilha},
  {Damke}, {Darling}, {Davidson}, {Dawson}, {de Icaza Lizaola}, {de la
  Macorra}, {de la Torre}, {De Lee}, {de Sainte Agathe}, {Deconto Machado},
  {Dell'Agli}, {Delubac}, {Diamond-Stanic}, {Donor}, {Downes}, {Drory}, {du Mas
  des Bourboux}, {Duckworth}, {Dwelly}, {Dyer}, {Ebelke}, {Davis Eigenbrot},
  {Eisenstein}, {Elsworth}, {Emsellem}, {Eracleous}, {Erfanianfar},
  {Escoffier}, {Fan}, {Fern{\'a}ndez Alvar}, {Fernandez-Trincado}, {Fernando
  Cirolini}, {Feuillet}, {Finoguenov}, {Fleming}, {Font-Ribera}, {Freischlad},
  {Frinchaboy}, {Fu}, {G{\'o}mez Maqueo Chew}, {Galbany}, {Garc{\'\i}a
  P{\'e}rez}, {Garcia-Dias}, {Garc{\'\i}a-Hern{\'a}ndez}, {Garma Oehmichen},
  {Gaulme}, {Gelfand}, {Gil-Mar{\'\i}n}, {Gillespie}, {Goddard}, {Gonz{\'a}lez
  Hern{\'a}ndez}, {Gonzalez-Perez}, {Grabowski}, {Green}, {Grier}, {Gueguen},
  {Guo}, {Guy}, {Hagen}, {Hall}, {Harding}, {Hasselquist}, {Hawley}, {Hayes},
  {Hearty}, {Hekker}, {Hernandez}, {Hernandez Toledo}, {Hogg},
  {Holley-Bockelmann}, {Holtzman}, {Hou}, {Hsieh}, {Hunt}, {Hutchinson},
  {Hwang}, {Jimenez Angel}, {Johnson}, {Jones}, {J{\"o}nsson}, {Jullo}, {Khan},
  {Kinemuchi}, {Kirkby}, {Kirkpatrick}, {Kitaura}, {Knapp}, {Kneib},
  {Kollmeier}, {Lacerna}, {Lane}, {Lang}, {Law}, {Le Goff}, {Lee}, {Li}, {Li},
  {Lian}, {Liang}, {Lima}, {Lin}, {Long}, {Lucatello}, {Lundgren}, {Mackereth},
  {MacLeod}, {Mahadevan}, {Maia}, {Majewski}, {Manchado}, {Maraston},
  {Mariappan}, {Marques-Chaves}, {Masseron}, {Masters}, {McDermid}, {McGreer},
  {Melendez}, {Meneses-Goytia}, {Merloni}, {Merrifield}, {Meszaros}, {Meza},
  {Minchev}, {Minniti}, {Mueller}, {Muller-Sanchez}, {Muna}, {Mu{\~n}oz},
  {Myers}, {Nair}, {Nandra}, {Ness}, {Newman}, {Nichol}, {Nidever},
  {Nitschelm}, {Noterdaeme}, {O'Connell}, {Oelkers}, {Oravetz}, {Oravetz},
  {Ort{\'\i}z}, {Osorio}, {Pace}, {Padilla}, {Palanque-Delabrouille},
  {Palicio}, {Pan}, {Pan}, {Parikh}, {P{\^a}ris}, {Park}, {Peirani},
  {Pellejero-Ibanez}, {Penny}, {Percival}, {Perez-Fournon}, {Petitjean},
  {Pieri}, {Pinsonneault}, {Pisani}, {Prada}, {Prakash}, {Queiroz}, {Raddick},
  {Raichoor}, {Barboza Rembold}, {Richstein}, {Riffel}, {Riffel}, {Rix},
  {Robin}, {Rodr{\'\i}guez Torres}, {Rom{\'a}n-Z{\'u}{\~n}iga}, {Ross},
  {Rossi}, {Ruan}, {Ruggeri}, {Ruiz}, {Salvato}, {S{\'a}nchez}, {S{\'a}nchez},
  {Sanchez Almeida}, {S{\'a}nchez-Gallego}, {Santana Rojas}, {Santiago},
  {Schiavon}, {Schimoia}, {Schlafly}, {Schlegel}, {Schneider}, {Schuster},
  {Schwope}, {Seo}, {Serenelli}, {Shen}, {Shen}, {Shetrone}, {Shull}, {Silva
  Aguirre}, {Simon}, {Skrutskie}, {Slosar}, {Smethurst}, {Smith}, {Sobeck},
  {Somers}, {Souter}, {Souto}, {Spindler}, {Stark}, {Stassun}, {Steinmetz},
  {Stello}, {Storchi-Bergmann}, {Streblyanska}, {Stringfellow}, {Su{\'a}rez},
  {Sun}, {Szigeti}, {Taghizadeh-Popp}, {Talbot}, {Tang}, {Tao}, {Tayar},
  {Tembe}, {Teske}, {Thakar}, {Thomas}, {Tissera}, {Tojeiro}, {Tremonti},
  {Troup}, {Urry}, {Valenzuela}, {van den Bosch}, {Vargas-Gonz{\'a}lez},
  {Vargas-Maga{\~n}a}, {Vazquez}, {Villanova}, {Vogt}, {Wake}, {Wang},
  {Weaver}, {Weijmans}, {Weinberg}, {Westfall}, {Whelan}, {Wilcots}, {Wild},
  {Williams}, {Wilson}, {Wood-Vasey}, {Wylezalek}, {Xiao}, {Yan}, {Yang},
  {Ybarra}, {Y{\`e}che}, {Zakamska}, {Zamora}, {Zarrouk}, {Zasowski}, {Zhang},
  {Zhao}, {Zhao}, {Zheng}, {Zheng}, {Zhou}, {Zhu}, {Zinn}, \&
  {Zou}}]{Abolfathi:2018vl}
{Abolfathi}, B., {Aguado}, D.~S., {Aguilar}, G., {et~al.} 2018, \apjs, 235, 42.
\newblock \doarXiv{1707.09322}

\bibitem[{{Astropy Collaboration} {et~al.}(2013){Astropy Collaboration},
  {Robitaille}, {Tollerud}, {Greenfield}, {Droettboom}, {Bray}, {Aldcroft},
  {Davis}, {Ginsburg}, {Price-Whelan}, {Kerzendorf}, {Conley}, {Crighton},
  {Barbary}, {Muna}, {Ferguson}, {Grollier}, {Parikh}, {Nair}, {Unther},
  {Deil}, {Woillez}, {Conseil}, {Kramer}, {Turner}, {Singer}, {Fox}, {Weaver},
  {Zabalza}, {Edwards}, {Azalee Bostroem}, {Burke}, {Casey}, {Crawford},
  {Dencheva}, {Ely}, {Jenness}, {Labrie}, {Lim}, {Pierfederici}, {Pontzen},
  {Ptak}, {Refsdal}, {Servillat}, \&
  {Streicher}}]{Astropy-Collaboration:2013uv}
{Astropy Collaboration}, {Robitaille}, T.~P., {Tollerud}, E.~J., {et~al.} 2013,
  \aap, 558, A33, \dodoi{10.1051/0004-6361/201322068}

\bibitem[{{Astropy Collaboration} {et~al.}(2018){Astropy Collaboration},
  {Price-Whelan}, {Sip{\H{o}}cz}, {G{\"u}nther}, {Lim}, {Crawford}, {Conseil},
  {Shupe}, {Craig}, {Dencheva}, {Ginsburg}, {Vand erPlas}, {Bradley},
  {P{\'e}rez-Su{\'a}rez}, {de Val-Borro}, {Aldcroft}, {Cruz}, {Robitaille},
  {Tollerud}, {Ardelean}, {Babej}, {Bach}, {Bachetti}, {Bakanov}, {Bamford},
  {Barentsen}, {Barmby}, {Baumbach}, {Berry}, {Biscani}, {Boquien}, {Bostroem},
  {Bouma}, {Brammer}, {Bray}, {Breytenbach}, {Buddelmeijer}, {Burke},
  {Calderone}, {Cano Rodr{\'\i}guez}, {Cara}, {Cardoso}, {Cheedella}, {Copin},
  {Corrales}, {Crichton}, {D'Avella}, {Deil}, {Depagne}, {Dietrich}, {Donath},
  {Droettboom}, {Earl}, {Erben}, {Fabbro}, {Ferreira}, {Finethy}, {Fox},
  {Garrison}, {Gibbons}, {Goldstein}, {Gommers}, {Greco}, {Greenfield},
  {Groener}, {Grollier}, {Hagen}, {Hirst}, {Homeier}, {Horton}, {Hosseinzadeh},
  {Hu}, {Hunkeler}, {Ivezi{\'c}}, {Jain}, {Jenness}, {Kanarek}, {Kendrew},
  {Kern}, {Kerzendorf}, {Khvalko}, {King}, {Kirkby}, {Kulkarni}, {Kumar},
  {Lee}, {Lenz}, {Littlefair}, {Ma}, {Macleod}, {Mastropietro}, {McCully},
  {Montagnac}, {Morris}, {Mueller}, {Mumford}, {Muna}, {Murphy}, {Nelson},
  {Nguyen}, {Ninan}, {N{\"o}the}, {Ogaz}, {Oh}, {Parejko}, {Parley}, {Pascual},
  {Patil}, {Patil}, {Plunkett}, {Prochaska}, {Rastogi}, {Reddy Janga},
  {Sabater}, {Sakurikar}, {Seifert}, {Sherbert}, {Sherwood-Taylor}, {Shih},
  {Sick}, {Silbiger}, {Singanamalla}, {Singer}, {Sladen}, {Sooley},
  {Sornarajah}, {Streicher}, {Teuben}, {Thomas}, {Tremblay}, {Turner},
  {Terr{\'o}n}, {van Kerkwijk}, {de la Vega}, {Watkins}, {Weaver}, {Whitmore},
  {Woillez}, {Zabalza}, \& {Astropy
  Contributors}}]{Astropy-Collaboration:2018vm}
{Astropy Collaboration}, {Price-Whelan}, A.~M., {Sip{\H{o}}cz}, B.~M., {et~al.}
  2018, \aj, 156, 123, \dodoi{10.3847/1538-3881/aabc4f}

\bibitem[{{Bahcall} \& {Peebles}(1969)}]{Bahcall:1969tt}
{Bahcall}, J.~N., \& {Peebles}, P.~J.~E. 1969, \apjl, 156, L7

\bibitem[{{Behroozi} {et~al.}(2019){Behroozi}, {Wechsler}, {Hearin}, \&
  {Conroy}}]{Behroozi:2019up}
{Behroozi}, P., {Wechsler}, R.~H., {Hearin}, A.~P., \& {Conroy}, C. 2019,
  \mnras, 488, 3143.
\newblock \doarXiv{1806.07893}

\bibitem[{{Berg} {et~al.}(2019){Berg}, {Howk}, {Lehner}, {Wotta}, {O'Meara},
  {Bowen}, {Burchett}, {Peeples}, \& {Tejos}}]{Berg:2019wq}
{Berg}, M.~A., {Howk}, J.~C., {Lehner}, N., {et~al.} 2019, \apj, 883, 5.
\newblock \doarXiv{1811.10717}

\bibitem[{{Bielby} {et~al.}(2019){Bielby}, {Stott}, {Cullen}, {Tripp},
  {Burchett}, {Fumagalli}, {Morris}, {Tejos}, {Crain}, {Bower}, \&
  {Prochaska}}]{Bielby:2019vv}
{Bielby}, R.~M., {Stott}, J.~P., {Cullen}, F., {et~al.} 2019, \mnras, 486, 21.
\newblock \doarXiv{1809.05544}

\bibitem[{{Bland-Hawthorn} \& {Gerhard}(2016)}]{Bland-Hawthorn:2016vl}
{Bland-Hawthorn}, J., \& {Gerhard}, O. 2016, \araa, 54, 529.
\newblock \doarXiv{1602.07702}

\bibitem[{{Boquien} {et~al.}(2019){Boquien}, {Burgarella}, {Roehlly}, {Buat},
  {Ciesla}, {Corre}, {Inoue}, \& {Salas}}]{Boquien:2019wx}
{Boquien}, M., {Burgarella}, D., {Roehlly}, Y., {et~al.} 2019, \aap, 622, A103.
\newblock \doarXiv{1811.03094}

\bibitem[{{Bordoloi} {et~al.}(2017){Bordoloi}, {Wagner}, {Heckman}, \&
  {Norman}}]{Bordoloi:2017vr}
{Bordoloi}, R., {Wagner}, A.~Y., {Heckman}, T.~M., \& {Norman}, C.~A. 2017,
  \apj, 848, 122.
\newblock \doarXiv{1605.07187}

\bibitem[{{Bordoloi} {et~al.}(2014){Bordoloi}, {Tumlinson}, {Werk},
  {Oppenheimer}, {Peeples}, {Prochaska}, {Tripp}, {Katz}, {Dav{\'e}}, {Fox},
  {Thom}, {Ford}, {Weinberg}, {Burchett}, \& {Kollmeier}}]{Bordoloi:2014uz}
{Bordoloi}, R., {Tumlinson}, J., {Werk}, J.~K., {et~al.} 2014, \apj, 796, 136.
\newblock \doarXiv{1406.0509}

\bibitem[{{Borkowski} {et~al.}(1990){Borkowski}, {Balbus}, \&
  {Fristrom}}]{Borkowski:1990ut}
{Borkowski}, K.~J., {Balbus}, S.~A., \& {Fristrom}, C.~C. 1990, \apj, 355, 501

\bibitem[{{Bruzual} \& {Charlot}(2003)}]{Bruzual:2003wd}
{Bruzual}, G., \& {Charlot}, S. 2003, \mnras, 344, 1000.
\newblock \doarXiv{astro-ph/0309134}

\bibitem[{{Bryan} \& {Norman}(1998)}]{Bryan:1998tw}
{Bryan}, G.~L., \& {Norman}, M.~L. 1998, \apj, 495, 80.
\newblock \doarXiv{astro-ph/9710107}

\bibitem[{{Burchett} {et~al.}(2018){Burchett}, {Tripp}, {Wang}, {Willmer},
  {Bowen}, \& {Jenkins}}]{Burchett:2018tn}
{Burchett}, J.~N., {Tripp}, T.~M., {Wang}, Q.~D., {et~al.} 2018, \mnras, 475,
  2067.
\newblock \doarXiv{1705.05892}

\bibitem[{{Burchett} {et~al.}(2019){Burchett}, {Tripp}, {Prochaska}, {Werk},
  {Tumlinson}, {Howk}, {Willmer}, {Lehner}, {Meiring}, {Bowen}, {Bordoloi},
  {Peeples}, {Jenkins}, {O'Meara}, {Tejos}, \& {Katz}}]{Burchett:2019vk}
{Burchett}, J.~N., {Tripp}, T.~M., {Prochaska}, J.~X., {et~al.} 2019, \apjl,
  877, L20.
\newblock \doarXiv{1810.06560}

\bibitem[{{Chabrier}(2003)}]{Chabrier:2003ta}
{Chabrier}, G. 2003, \pasp, 115, 763.
\newblock \doarXiv{astro-ph/0304382}

\bibitem[{{Chambers} {et~al.}(2016){Chambers}, {Magnier}, {Metcalfe},
  {Flewelling}, {Huber}, {Waters}, {Denneau}, {Draper}, {Farrow}, {Finkbeiner},
  {Holmberg}, {Koppenhoefer}, {Price}, {Rest}, {Saglia}, {Schlafly}, {Smartt},
  {Sweeney}, {Wainscoat}, {Burgett}, {Chastel}, {Grav}, {Heasley}, {Hodapp},
  {Jedicke}, {Kaiser}, {Kudritzki}, {Luppino}, {Lupton}, {Monet}, {Morgan},
  {Onaka}, {Shiao}, {Stubbs}, {Tonry}, {White}, {Ba{\~n}ados}, {Bell},
  {Bender}, {Bernard}, {Boegner}, {Boffi}, {Botticella}, {Calamida},
  {Casertano}, {Chen}, {Chen}, {Cole}, {Deacon}, {Frenk}, {Fitzsimmons},
  {Gezari}, {Gibbs}, {Goessl}, {Goggia}, {Gourgue}, {Goldman}, {Grant},
  {Grebel}, {Hambly}, {Hasinger}, {Heavens}, {Heckman}, {Henderson}, {Henning},
  {Holman}, {Hopp}, {Ip}, {Isani}, {Jackson}, {Keyes}, {Koekemoer}, {Kotak},
  {Le}, {Liska}, {Long}, {Lucey}, {Liu}, {Martin}, {Masci}, {McLean}, {Mindel},
  {Misra}, {Morganson}, {Murphy}, {Obaika}, {Narayan}, {Nieto-Santisteban},
  {Norberg}, {Peacock}, {Pier}, {Postman}, {Primak}, {Rae}, {Rai}, {Riess},
  {Riffeser}, {Rix}, {R{\"o}ser}, {Russel}, {Rutz}, {Schilbach}, {Schultz},
  {Scolnic}, {Strolger}, {Szalay}, {Seitz}, {Small}, {Smith}, {Soderblom},
  {Taylor}, {Thomson}, {Taylor}, {Thakar}, {Thiel}, {Thilker}, {Unger},
  {Urata}, {Valenti}, {Wagner}, {Walder}, {Walter}, {Watters}, {Werner},
  {Wood-Vasey}, \& {Wyse}}]{Chambers:2016vk}
{Chambers}, K.~C., {Magnier}, E.~A., {Metcalfe}, N., {et~al.} 2016, arXiv
  e-prints, arXiv:1612.05560.
\newblock \doarXiv{1612.05560}

\bibitem[{{Chen} {et~al.}(2018){Chen}, {Zahedy}, {Johnson}, {Pierce}, {Huang},
  {Weiner}, \& {Gauthier}}]{Chen:2018tu}
{Chen}, H.-W., {Zahedy}, F.~S., {Johnson}, S.~D., {et~al.} 2018, \mnras, 479,
  2547.
\newblock \doarXiv{1805.07364}

\bibitem[{{Cutri} {et~al.}(2013){Cutri}, {Wright}, {Conrow}, {Fowler},
  {Eisenhardt}, {Grillmair}, {Kirkpatrick}, {Masci}, {McCallon}, {Wheelock},
  {Fajardo-Acosta}, {Yan}, {Benford}, {Harbut}, {Jarrett}, {Lake}, {Leisawitz},
  {Ressler}, {Stanford}, {Tsai}, {Liu}, {Helou}, {Mainzer}, {Gettings},
  {Gonzalez}, {Hoffman}, {Marsh}, {Padgett}, {Skrutskie}, {Beck}, {Papin}, \&
  {Wittman}}]{Cutri:2013tn}
{Cutri}, R.~M., {Wright}, E.~L., {Conrow}, T., {et~al.} 2013, {Explanatory
  Supplement to the AllWISE Data Release Products}, Tech. rep.

\bibitem[{{Danforth} {et~al.}(2016){Danforth}, {Keeney}, {Tilton}, {Shull},
  {Stocke}, {Stevans}, {Pieri}, {Savage}, {France}, {Syphers}, {Smith},
  {Green}, {Froning}, {Penton}, \& {Osterman}}]{Danforth:2016uj}
{Danforth}, C.~W., {Keeney}, B.~A., {Tilton}, E.~M., {et~al.} 2016, \apj, 817,
  111.
\newblock \doarXiv{1402.2655}

\bibitem[{{Davies} {et~al.}(2020){Davies}, {Crain}, {Oppenheimer}, \&
  {Schaye}}]{Davies:2020wd}
{Davies}, J.~J., {Crain}, R.~A., {Oppenheimer}, B.~D., \& {Schaye}, J. 2020,
  \mnras, 491, 4462.
\newblock \doarXiv{1908.11380}

\bibitem[{{Dey} {et~al.}(2019){Dey}, {Schlegel}, {Lang}, {Blum}, {Burleigh},
  {Fan}, {Findlay}, {Finkbeiner}, {Herrera}, {Juneau}, {Landriau}, {Levi},
  {McGreer}, {Meisner}, {Myers}, {Moustakas}, {Nugent}, {Patej}, {Schlafly},
  {Walker}, {Valdes}, {Weaver}, {Y{\`e}che}, {Zou}, {Zhou}, {Abareshi},
  {Abbott}, {Abolfathi}, {Aguilera}, {Alam}, {Allen}, {Alvarez}, {Annis},
  {Ansarinejad}, {Aubert}, {Beechert}, {Bell}, {BenZvi}, {Beutler}, {Bielby},
  {Bolton}, {Brice{\~n}o}, {Buckley-Geer}, {Butler}, {Calamida}, {Carlberg},
  {Carter}, {Casas}, {Castander}, {Choi}, {Comparat}, {Cukanovaite}, {Delubac},
  {DeVries}, {Dey}, {Dhungana}, {Dickinson}, {Ding}, {Donaldson}, {Duan},
  {Duckworth}, {Eftekharzadeh}, {Eisenstein}, {Etourneau}, {Fagrelius},
  {Farihi}, {Fitzpatrick}, {Font-Ribera}, {Fulmer}, {G{\"a}nsicke},
  {Gaztanaga}, {George}, {Gerdes}, {Gontcho}, {Gorgoni}, {Green}, {Guy},
  {Harmer}, {Hernandez}, {Honscheid}, {Huang}, {James}, {Jannuzi}, {Jiang},
  {Joyce}, {Karcher}, {Karkar}, {Kehoe}, {Kneib}, {Kueter-Young}, {Lan},
  {Lauer}, {Le Guillou}, {Le Van Suu}, {Lee}, {Lesser}, {Perreault Levasseur},
  {Li}, {Mann}, {Marshall}, {Mart{\'\i}nez-V{\'a}zquez}, {Martini}, {du Mas des
  Bourboux}, {McManus}, {Meier}, {M{\'e}nard}, {Metcalfe},
  {Mu{\~n}oz-Guti{\'e}rrez}, {Najita}, {Napier}, {Narayan}, {Newman}, {Nie},
  {Nord}, {Norman}, {Olsen}, {Paat}, {Palanque-Delabrouille}, {Peng},
  {Poppett}, {Poremba}, {Prakash}, {Rabinowitz}, {Raichoor}, {Rezaie},
  {Robertson}, {Roe}, {Ross}, {Ross}, {Rudnick}, {Safonova}, {Saha},
  {S{\'a}nchez}, {Savary}, {Schweiker}, {Scott}, {Seo}, {Shan}, {Silva},
  {Slepian}, {Soto}, {Sprayberry}, {Staten}, {Stillman}, {Stupak}, {Summers},
  {Sien Tie}, {Tirado}, {Vargas-Maga{\~n}a}, {Vivas}, {Wechsler}, {Williams},
  {Yang}, {Yang}, {Yapici}, {Zaritsky}, {Zenteno}, {Zhang}, {Zhang}, {Zhou}, \&
  {Zhou}}]{Dey:2019uk}
{Dey}, A., {Schlegel}, D.~J., {Lang}, D., {et~al.} 2019, \aj, 157, 168.
\newblock \doarXiv{1804.08657}

\bibitem[{{Dopita} \& {Sutherland}(1996)}]{Dopita:1996wz}
{Dopita}, M.~A., \& {Sutherland}, R.~S. 1996, \apjs, 102, 161

\bibitem[{{Dutta} {et~al.}(2021){Dutta}, {Fumagalli}, {Fossati}, {Bielby},
  {Stott}, {Lofthouse}, {Cantalupo}, {Cullen}, {Crain}, {Tripp}, {Prochaska},
  {Battaia}, {Burchett}, {Fynbo}, {Murphy}, {Schaye}, {Tejos}, \&
  {Theuns}}]{Dutta:2021vq}
{Dutta}, R., {Fumagalli}, M., {Fossati}, M., {et~al.} 2021, \mnras.
\newblock \doarXiv{2109.10927}

\bibitem[{{Edgar} \& {Chevalier}(1986)}]{Edgar:1986tr}
{Edgar}, R.~J., \& {Chevalier}, R.~A. 1986, \apjl, 310, L27

\bibitem[{{Faerman} {et~al.}(2020){Faerman}, {Sternberg}, \&
  {McKee}}]{Faerman:2020wc}
{Faerman}, Y., {Sternberg}, A., \& {McKee}, C.~F. 2020, \apj, 893, 82.
\newblock \doarXiv{1909.09169}

\bibitem[{{Faucher-Gigu{\`e}re}(2020)}]{Faucher-Giguere:2020wt}
{Faucher-Gigu{\`e}re}, C.-A. 2020, \mnras, 493, 1614.
\newblock \doarXiv{1903.08657}

\bibitem[{{Fielding} {et~al.}(2020){Fielding}, {Tonnesen}, {DeFelippis}, {Li},
  {Su}, {Bryan}, {Kim}, {Forbes}, {Somerville}, {Battaglia}, {Schneider}, {Li},
  {Choi}, {Hayward}, \& {Hernquist}}]{Fielding:2020vf}
{Fielding}, D.~B., {Tonnesen}, S., {DeFelippis}, D., {et~al.} 2020, \apj, 903,
  32.
\newblock \doarXiv{2006.16316}

\bibitem[{{Finn} {et~al.}(2016){Finn}, {Morris}, {Tejos}, {Crighton}, {Perry},
  {Fumagalli}, {Bielby}, {Theuns}, {Schaye}, {Shanks}, {Liske}, {Gunawardhana},
  \& {Bartle}}]{Finn:2016tq}
{Finn}, C.~W., {Morris}, S.~L., {Tejos}, N., {et~al.} 2016, \mnras, 460, 590.
\newblock \doarXiv{1604.02150}

\bibitem[{{Gimeno} {et~al.}(2016){Gimeno}, {Roth}, {Chiboucas}, {Hibon},
  {Boucher}, {White}, {Rippa}, {Labrie}, {Turner}, {Hanna}, {Lazo},
  {P{\'e}rez}, {Rogers}, {Rojas}, {Placco}, \& {Murowinski}}]{Gimeno:2016ul}
{Gimeno}, G., {Roth}, K., {Chiboucas}, K., {et~al.} 2016, in Society of
  Photo-Optical Instrumentation Engineers (SPIE) Conference Series, Vol. 9908,
  Ground-based and Airborne Instrumentation for Astronomy VI, ed. C.~J.
  {Evans}, L.~{Simard}, \& H.~{Takami}, 99082S

\bibitem[{{Gnat} \& {Sternberg}(2007)}]{Gnat:2007tt}
{Gnat}, O., \& {Sternberg}, A. 2007, \apjs, 168, 213.
\newblock \doarXiv{astro-ph/0608181}

\bibitem[{{Gnat} {et~al.}(2010){Gnat}, {Sternberg}, \& {McKee}}]{Gnat:2010wu}
{Gnat}, O., {Sternberg}, A., \& {McKee}, C.~F. 2010, \apj, 718, 1315.
\newblock \doarXiv{1002.1309}

\bibitem[{{Green} {et~al.}(2012){Green}, {Froning}, {Osterman}, {Ebbets},
  {Heap}, {Leitherer}, {Linsky}, {Savage}, {Sembach}, {Shull}, {Siegmund},
  {Snow}, {Spencer}, {Stern}, {Stocke}, {Welsh}, {B{\'e}land}, {Burgh},
  {Danforth}, {France}, {Keeney}, {McPhate}, {Penton}, {Andrews},
  {Brownsberger}, {Morse}, \& {Wilkinson}}]{Green:2012vw}
{Green}, J.~C., {Froning}, C.~S., {Osterman}, S., {et~al.} 2012, \apj, 744, 60.
\newblock \doarXiv{1110.0462}

\bibitem[{{Haardt} \& {Madau}(2012)}]{Haardt:2012ti}
{Haardt}, F., \& {Madau}, P. 2012, \apj, 746, 125.
\newblock \doarXiv{1105.2039}

\bibitem[{Harris {et~al.}(2020)Harris, Millman, van~der Walt, Gommers,
  Virtanen, Cournapeau, Wieser, Taylor, Berg, Smith, Kern, Picus, Hoyer, van
  Kerkwijk, Brett, Haldane, del R{\'{i}}o, Wiebe, Peterson,
  G{\'{e}}rard-Marchant, Sheppard, Reddy, Weckesser, Abbasi, Gohlke, \&
  Oliphant}]{Harris:2020ti}
Harris, C.~R., Millman, K.~J., van~der Walt, S.~J., {et~al.} 2020, Nature, 585,
  357, \dodoi{10.1038/s41586-020-2649-2}

\bibitem[{{Heckman} {et~al.}(2002){Heckman}, {Norman}, {Strickland}, \&
  {Sembach}}]{Heckman:2002wz}
{Heckman}, T.~M., {Norman}, C.~A., {Strickland}, D.~K., \& {Sembach}, K.~R.
  2002, \apj, 577, 691.
\newblock \doarXiv{astro-ph/0205556}

\bibitem[{{Ho} {et~al.}(2021){Ho}, {Martin}, \& {Schaye}}]{Ho:2021we}
{Ho}, S.~H., {Martin}, C.~L., \& {Schaye}, J. 2021, arXiv e-prints,
  arXiv:2110.01633.
\newblock \doarXiv{2110.01633}

\bibitem[{{Hogg}(1999)}]{Hogg:1999tj}
{Hogg}, D.~W. 1999, arXiv e-prints, astro.
\newblock \doarXiv{astro-ph/9905116}

\bibitem[{{Hook} {et~al.}(2004){Hook}, {J{\o}rgensen}, {Allington-Smith},
  {Davies}, {Metcalfe}, {Murowinski}, \& {Crampton}}]{Hook:2004wu}
{Hook}, I.~M., {J{\o}rgensen}, I., {Allington-Smith}, J.~R., {et~al.} 2004,
  \pasp, 116, 425

\bibitem[{{Hu} \& {Kravtsov}(2003)}]{Hu:2003tf}
{Hu}, W., \& {Kravtsov}, A.~V. 2003, \apj, 584, 702.
\newblock \doarXiv{astro-ph/0203169}

\bibitem[{{Hunter}(2007)}]{Hunter:2007ux}
{Hunter}, J.~D. 2007, Computing in Science and Engineering, 9, 90,
  \dodoi{10.1109/MCSE.2007.55}

\bibitem[{{Johnson} {et~al.}(2015){Johnson}, {Chen}, \&
  {Mulchaey}}]{Johnson:2015tj}
{Johnson}, S.~D., {Chen}, H.-W., \& {Mulchaey}, J.~S. 2015, \mnras, 449, 3263.
\newblock \doarXiv{1503.04199}

\bibitem[{{Johnson} {et~al.}(2017){Johnson}, {Chen}, {Mulchaey}, {Schaye}, \&
  {Straka}}]{Johnson:2017tp}
{Johnson}, S.~D., {Chen}, H.-W., {Mulchaey}, J.~S., {Schaye}, J., \& {Straka},
  L.~A. 2017, \apjl, 850, L10.
\newblock \doarXiv{1710.06441}

\bibitem[{{Keeney} {et~al.}(2017){Keeney}, {Stocke}, {Danforth}, {Shull},
  {Pratt}, {Froning}, {Green}, {Penton}, \& {Savage}}]{Keeney:2017wk}
{Keeney}, B.~A., {Stocke}, J.~T., {Danforth}, C.~W., {et~al.} 2017, \apjs, 230,
  6.
\newblock \doarXiv{1704.00235}

\bibitem[{{Keeney} {et~al.}(2018){Keeney}, {Stocke}, {Pratt}, {Davis},
  {Syphers}, {Danforth}, {Shull}, {Froning}, {Green}, {Penton}, \&
  {Savage}}]{Keeney:2018wp}
{Keeney}, B.~A., {Stocke}, J.~T., {Pratt}, C.~T., {et~al.} 2018, \apjs, 237,
  11.
\newblock \doarXiv{1805.08767}

\bibitem[{{Keller} {et~al.}(2020){Keller}, {Kruijssen}, \&
  {Wadsley}}]{Keller:2020we}
{Keller}, B.~W., {Kruijssen}, J.~M.~D., \& {Wadsley}, J.~W. 2020, \mnras, 493,
  2149.
\newblock \doarXiv{1909.00815}

\bibitem[{{Lehner} {et~al.}(2020){Lehner}, {Berek}, {Howk}, {Wakker},
  {Tumlinson}, {Jenkins}, {Prochaska}, {Augustin}, {Ji}, {Faucher-Gigu{\`e}re},
  {Hafen}, {Peeples}, {Barger}, {Berg}, {Bordoloi}, {Brown}, {Fox}, {Gilbert},
  {Guhathakurta}, {Kalirai}, {Lockman}, {O'Meara}, {Pisano}, {Ribaudo}, \&
  {Werk}}]{Lehner:2020tk}
{Lehner}, N., {Berek}, S.~C., {Howk}, J.~C., {et~al.} 2020, \apj, 900, 9.
\newblock \doarXiv{2002.07818}

\bibitem[{{Li} \& {Tonnesen}(2020)}]{Li:2020wn}
{Li}, M., \& {Tonnesen}, S. 2020, \apj, 898, 148.
\newblock \doarXiv{1910.14235}

\bibitem[{{Lochhaas} {et~al.}(2021){Lochhaas}, {Tumlinson}, {O'Shea},
  {Peeples}, {Smith}, {Werk}, {Augustin}, \& {Simons}}]{Lochhaas:2021to}
{Lochhaas}, C., {Tumlinson}, J., {O'Shea}, B.~W., {et~al.} 2021, arXiv
  e-prints, arXiv:2102.08393.
\newblock \doarXiv{2102.08393}

\bibitem[{{Mathews} \& {Prochaska}(2017)}]{Mathews:2017uf}
{Mathews}, W.~G., \& {Prochaska}, J.~X. 2017, \apjl, 846, L24.
\newblock \doarXiv{1708.07140}

\bibitem[{{McQuinn} \& {Werk}(2018)}]{McQuinn:2018vk}
{McQuinn}, M., \& {Werk}, J.~K. 2018, \apj, 852, 33.
\newblock \doarXiv{1703.03422}

\bibitem[{{Mina} {et~al.}(2020){Mina}, {Shen}, {Keller}, {Mayer}, {Madau}, \&
  {Wadsley}}]{Mina:2020tx}
{Mina}, M., {Shen}, S., {Keller}, B.~W., {et~al.} 2020, arXiv e-prints,
  arXiv:2009.06646.
\newblock \doarXiv{2009.06646}

\bibitem[{{Moos} {et~al.}(2000){Moos}, {Cash}, {Cowie}, {Davidsen}, {Dupree},
  {Feldman}, {Friedman}, {Green}, {Green}, {Gry}, {Hutchings}, {Jenkins},
  {Linsky}, {Malina}, {Michalitsianos}, {Savage}, {Shull}, {Siegmund}, {Snow},
  {Sonneborn}, {Vidal-Madjar}, {Willis}, {Woodgate}, {York}, {Ake},
  {Andersson}, {Andrews}, {Barkhouser}, {Bianchi}, {Blair}, {Brownsberger},
  {Cha}, {Chayer}, {Conard}, {Fullerton}, {Gaines}, {Grange}, {Gummin},
  {Hebrard}, {Kriss}, {Kruk}, {Mark}, {McCarthy}, {Morbey}, {Murowinski},
  {Murphy}, {Oegerle}, {Ohl}, {Oliveira}, {Osterman}, {Sahnow}, {Saisse},
  {Sembach}, {Weaver}, {Welsh}, {Wilkinson}, \& {Zheng}}]{Moos:2000vf}
{Moos}, H.~W., {Cash}, W.~C., {Cowie}, L.~L., {et~al.} 2000, \apjl, 538, L1.
\newblock \doarXiv{astro-ph/0005529}

\bibitem[{{Moos} {et~al.}(2002){Moos}, {Sembach}, {Vidal-Madjar}, {York},
  {Friedman}, {H{\'e}brard}, {Kruk}, {Lehner}, {Lemoine}, {Sonneborn}, {Wood},
  {Ake}, {Andr{\'e}}, {Blair}, {Chayer}, {Gry}, {Dupree}, {Ferlet}, {Feldman},
  {Green}, {Howk}, {Hutchings}, {Jenkins}, {Linsky}, {Murphy}, {Oegerle},
  {Oliveira}, {Roth}, {Sahnow}, {Savage}, {Shull}, {Tripp}, {Weiler}, {Welsh},
  {Wilkinson}, \& {Woodgate}}]{Moos:2002ux}
{Moos}, H.~W., {Sembach}, K.~R., {Vidal-Madjar}, A., {et~al.} 2002, \apjs, 140,
  3.
\newblock \doarXiv{astro-ph/0112519}

\bibitem[{{Nelson} {et~al.}(2018){Nelson}, {Kauffmann}, {Pillepich}, {Genel},
  {Springel}, {Pakmor}, {Hernquist}, {Weinberger}, {Torrey}, {Vogelsberger}, \&
  {Marinacci}}]{Nelson:2018wy}
{Nelson}, D., {Kauffmann}, G., {Pillepich}, A., {et~al.} 2018, \mnras, 477,
  450.
\newblock \doarXiv{1712.00016}

\bibitem[{{Oppenheimer} \& {Schaye}(2013)}]{Oppenheimer:2013vs}
{Oppenheimer}, B.~D., \& {Schaye}, J. 2013, \mnras, 434, 1043.
\newblock \doarXiv{1302.5710}

\bibitem[{{Oppenheimer} {et~al.}(2018){Oppenheimer}, {Schaye}, {Crain}, {Werk},
  \& {Richings}}]{Oppenheimer:2018vn}
{Oppenheimer}, B.~D., {Schaye}, J., {Crain}, R.~A., {Werk}, J.~K., \&
  {Richings}, A.~J. 2018, \mnras, 481, 835.
\newblock \doarXiv{1709.07577}

\bibitem[{{Oppenheimer} {et~al.}(2016){Oppenheimer}, {Crain}, {Schaye},
  {Rahmati}, {Richings}, {Trayford}, {Tumlinson}, {Bower}, {Schaller}, \&
  {Theuns}}]{Oppenheimer:2016wy}
{Oppenheimer}, B.~D., {Crain}, R.~A., {Schaye}, J., {et~al.} 2016, \mnras, 460,
  2157.
\newblock \doarXiv{1603.05984}

\bibitem[{{Oppenheimer} {et~al.}(2020){Oppenheimer}, {Davies}, {Crain},
  {Wijers}, {Schaye}, {Werk}, {Burchett}, {Trayford}, \&
  {Horton}}]{Oppenheimer:2020tx}
{Oppenheimer}, B.~D., {Davies}, J.~J., {Crain}, R.~A., {et~al.} 2020, \mnras,
  491, 2939.
\newblock \doarXiv{1904.05904}

\bibitem[{{Pandya} {et~al.}(2021){Pandya}, {Fielding},
  {Angl{\'e}s-Alc{\'a}zar}, {Somerville}, {Bryan}, {Hayward}, {Stern}, {Kim},
  {Quataert}, {Forbes}, {Faucher-Gigu{\`e}re}, {Feldmann}, {Hafen}, {Hopkins},
  {Kere{\v{s}}}, {Murray}, \& {Wetzel}}]{Pandya:2021vk}
{Pandya}, V., {Fielding}, D., {Angl{\'e}s-Alc{\'a}zar}, D., {et~al.} 2021,
  arXiv e-prints, arXiv:2103.06891.
\newblock \doarXiv{2103.06891}

\bibitem[{{Peeples} {et~al.}(2017){Peeples}, {Tumlinson}, {Fox}, {Aloisi},
  {Fleming}, {Jedrzejewski}, {Oliveira}, {Ayres}, {Danforth}, {Keeney}, \&
  {Jenkins}}]{Peeples:2017wd}
{Peeples}, M., {Tumlinson}, J., {Fox}, A., {et~al.} 2017, {The Hubble
  Spectroscopic Legacy Archive}, Tech. rep.

\bibitem[{{Planck Collaboration} {et~al.}(2016){Planck Collaboration}, {Ade},
  {Aghanim}, {Arnaud}, {Ashdown}, {Aumont}, {Baccigalupi}, {Banday},
  {Barreiro}, {Bartlett}, {Bartolo}, {Battaner}, {Battye}, {Benabed},
  {Beno{\^\i}t}, {Benoit-L{\'e}vy}, {Bernard}, {Bersanelli}, {Bielewicz},
  {Bock}, {Bonaldi}, {Bonavera}, {Bond}, {Borrill}, {Bouchet}, {Boulanger},
  {Bucher}, {Burigana}, {Butler}, {Calabrese}, {Cardoso}, {Catalano},
  {Challinor}, {Chamballu}, {Chary}, {Chiang}, {Chluba}, {Christensen},
  {Church}, {Clements}, {Colombi}, {Colombo}, {Combet}, {Coulais}, {Crill},
  {Curto}, {Cuttaia}, {Danese}, {Davies}, {Davis}, {de Bernardis}, {de Rosa},
  {de Zotti}, {Delabrouille}, {D{\'e}sert}, {Di Valentino}, {Dickinson},
  {Diego}, {Dolag}, {Dole}, {Donzelli}, {Dor{\'e}}, {Douspis}, {Ducout},
  {Dunkley}, {Dupac}, {Efstathiou}, {Elsner}, {En{\ss}lin}, {Eriksen},
  {Farhang}, {Fergusson}, {Finelli}, {Forni}, {Frailis}, {Fraisse},
  {Franceschi}, {Frejsel}, {Galeotta}, {Galli}, {Ganga}, {Gauthier}, {Gerbino},
  {Ghosh}, {Giard}, {Giraud-H{\'e}raud}, {Giusarma}, {Gjerl{\o}w},
  {Gonz{\'a}lez-Nuevo}, {G{\'o}rski}, {Gratton}, {Gregorio}, {Gruppuso},
  {Gudmundsson}, {Hamann}, {Hansen}, {Hanson}, {Harrison}, {Helou},
  {Henrot-Versill{\'e}}, {Hern{\'a}ndez-Monteagudo}, {Herranz}, {Hildebrandt},
  {Hivon}, {Hobson}, {Holmes}, {Hornstrup}, {Hovest}, {Huang}, {Huffenberger},
  {Hurier}, {Jaffe}, {Jaffe}, {Jones}, {Juvela}, {Keih{\"a}nen}, {Keskitalo},
  {Kisner}, {Kneissl}, {Knoche}, {Knox}, {Kunz}, {Kurki-Suonio}, {Lagache},
  {L{\"a}hteenm{\"a}ki}, {Lamarre}, {Lasenby}, {Lattanzi}, {Lawrence}, {Leahy},
  {Leonardi}, {Lesgourgues}, {Levrier}, {Lewis}, {Liguori}, {Lilje},
  {Linden-V{\o}rnle}, {L{\'o}pez-Caniego}, {Lubin}, {Mac{\'\i}as-P{\'e}rez},
  {Maggio}, {Maino}, {Mandolesi}, {Mangilli}, {Marchini}, {Maris}, {Martin},
  {Martinelli}, {Mart{\'\i}nez-Gonz{\'a}lez}, {Masi}, {Matarrese}, {McGehee},
  {Meinhold}, {Melchiorri}, {Melin}, {Mendes}, {Mennella}, {Migliaccio},
  {Millea}, {Mitra}, {Miville-Desch{\^e}nes}, {Moneti}, {Montier}, {Morgante},
  {Mortlock}, {Moss}, {Munshi}, {Murphy}, {Naselsky}, {Nati}, {Natoli},
  {Netterfield}, {N{\o}rgaard-Nielsen}, {Noviello}, {Novikov}, {Novikov},
  {Oxborrow}, {Paci}, {Pagano}, {Pajot}, {Paladini}, {Paoletti}, {Partridge},
  {Pasian}, {Patanchon}, {Pearson}, {Perdereau}, {Perotto}, {Perrotta},
  {Pettorino}, {Piacentini}, {Piat}, {Pierpaoli}, {Pietrobon}, {Plaszczynski},
  {Pointecouteau}, {Polenta}, {Popa}, {Pratt}, {Pr{\'e}zeau}, {Prunet},
  {Puget}, {Rachen}, {Reach}, {Rebolo}, {Reinecke}, {Remazeilles}, {Renault},
  {Renzi}, {Ristorcelli}, {Rocha}, {Rosset}, {Rossetti}, {Roudier},
  {Rouill{\'e} d'Orfeuil}, {Rowan-Robinson}, {Rubi{\~n}o-Mart{\'\i}n},
  {Rusholme}, {Said}, {Salvatelli}, {Salvati}, {Sandri}, {Santos},
  {Savelainen}, {Savini}, {Scott}, {Seiffert}, {Serra}, {Shellard}, {Spencer},
  {Spinelli}, {Stolyarov}, {Stompor}, {Sudiwala}, {Sunyaev}, {Sutton},
  {Suur-Uski}, {Sygnet}, {Tauber}, {Terenzi}, {Toffolatti}, {Tomasi},
  {Tristram}, {Trombetti}, {Tucci}, {Tuovinen}, {T{\"u}rler}, {Umana},
  {Valenziano}, {Valiviita}, {Van Tent}, {Vielva}, {Villa}, {Wade}, {Wandelt},
  {Wehus}, {White}, {White}, {Wilkinson}, {Yvon}, {Zacchei}, \&
  {Zonca}}]{Planck-Collaboration:2016tp}
{Planck Collaboration}, {Ade}, P.~A.~R., {Aghanim}, N., {et~al.} 2016, \aap,
  594, A13.
\newblock \doarXiv{1502.01589}

\bibitem[{{Pointon} {et~al.}(2017){Pointon}, {Nielsen}, {Kacprzak}, {Muzahid},
  {Churchill}, \& {Charlton}}]{Pointon:2017wl}
{Pointon}, S.~K., {Nielsen}, N.~M., {Kacprzak}, G.~G., {et~al.} 2017, \apj,
  844, 23.
\newblock \doarXiv{1706.03895}

\bibitem[{{Prochaska} {et~al.}(2011){Prochaska}, {Weiner}, {Chen}, {Mulchaey},
  \& {Cooksey}}]{Prochaska:2011vo}
{Prochaska}, J.~X., {Weiner}, B., {Chen}, H.~W., {Mulchaey}, J., \& {Cooksey},
  K. 2011, \apj, 740, 91.
\newblock \doarXiv{1103.1891}

\bibitem[{{Prochaska} {et~al.}(2017){Prochaska}, {Tejos}, {Crighton},
  {jnburchett}, {tiffanyhsyu}, {Tuo-Ji}, {marijana777}, {ktirimba}, {jhennawi},
  {Cooke}, {O'Meara}, \& {Werk}}]{Prochaska:2017vh}
{Prochaska}, J.~X., {Tejos}, N., {Crighton}, N., {et~al.} 2017,
  {Linetools/Linetools: Third Minor Release}, v0.3,  Zenodo,
  \dodoi{10.5281/zenodo.1036773}

\bibitem[{{Prochaska} {et~al.}(2019){Prochaska}, {Burchett}, {Tripp}, {Werk},
  {Willmer}, {Howk}, {Lange}, {Tejos}, {Meiring}, {Tumlinson}, {Lehner},
  {Ford}, \& {Dav{\'e}}}]{Prochaska:2019vy}
{Prochaska}, J.~X., {Burchett}, J.~N., {Tripp}, T.~M., {et~al.} 2019, \apjs,
  243, 24.
\newblock \doarXiv{1908.07675}

\bibitem[{{Qu} \& {Bregman}(2018)}]{Qu:2018va}
{Qu}, Z., \& {Bregman}, J.~N. 2018, \apj, 862, 23.
\newblock \doarXiv{1804.08784}

\bibitem[{{Roy} {et~al.}(2021){Roy}, {Nath}, \& {Voit}}]{Roy:2021vv}
{Roy}, M., {Nath}, B.~B., \& {Voit}, G.~M. 2021, \mnras, 507, 3849.
\newblock \doarXiv{2108.08320}

\bibitem[{Salvatier {et~al.}(2016)Salvatier, Wiecki, \&
  Fonnesbeck}]{Salvatier:2016ui}
Salvatier, J., Wiecki, T.~V., \& Fonnesbeck, C. 2016, {PeerJ} Computer Science,
  2, e55, \dodoi{10.7717/peerj-cs.55}

\bibitem[{{Salvatier} {et~al.}(2016){Salvatier}, {Wiecki{\^a}}, \&
  {Fonnesbeck}}]{Salvatier:2016ul}
{Salvatier}, J., {Wiecki{\^a}}, T.~V., \& {Fonnesbeck}, C. 2016, {PyMC3: Python
  probabilistic programming framework}, Astrophysics Source Code Library.
\newblock \doeprint{1610.016}

\bibitem[{{Skibba} {et~al.}(2012){Skibba}, {Engelbracht}, {Aniano}, {Babler},
  {Bernard}, {Bot}, {Carlson}, {Galametz}, {Galliano}, {Gordon}, {Hony},
  {Israel}, {Lebouteiller}, {Li}, {Madden}, {Meixner}, {Misselt}, {Montiel},
  {Okumura}, {Panuzzo}, {Paradis}, {Roman-Duval}, {Rubio}, {Sauvage}, {Seale},
  {Srinivasan}, \& {van Loon}}]{Skibba:2012uj}
{Skibba}, R.~A., {Engelbracht}, C.~W., {Aniano}, G., {et~al.} 2012, \apj, 761,
  42.
\newblock \doarXiv{1210.7812}

\bibitem[{{Slavin} {et~al.}(1993){Slavin}, {Shull}, \&
  {Begelman}}]{Slavin:1993ve}
{Slavin}, J.~D., {Shull}, J.~M., \& {Begelman}, M.~C. 1993, \apj, 407, 83

\bibitem[{{Stern} {et~al.}(2018){Stern}, {Faucher-Gigu{\`e}re}, {Hennawi},
  {Hafen}, {Johnson}, \& {Fielding}}]{Stern:2018to}
{Stern}, J., {Faucher-Gigu{\`e}re}, C.-A., {Hennawi}, J.~F., {et~al.} 2018,
  \apj, 865, 91.
\newblock \doarXiv{1803.05446}

\bibitem[{{Stern} {et~al.}(2019){Stern}, {Fielding}, {Faucher-Gigu{\`e}re}, \&
  {Quataert}}]{Stern:2019tp}
{Stern}, J., {Fielding}, D., {Faucher-Gigu{\`e}re}, C.-A., \& {Quataert}, E.
  2019, \mnras, 488, 2549.
\newblock \doarXiv{1906.07737}

\bibitem[{{Stocke} {et~al.}(2019){Stocke}, {Keeney}, {Danforth}, {Oppenheimer},
  {Pratt}, {Berlind}, {Impey}, \& {Jannuzi}}]{Stocke:2019vr}
{Stocke}, J.~T., {Keeney}, B.~A., {Danforth}, C.~W., {et~al.} 2019, \apjs, 240,
  15.
\newblock \doarXiv{1811.12374}

\bibitem[{{Strawn} {et~al.}(2021){Strawn}, {Roca-F{\`a}brega}, {Mandelker},
  {Primack}, {Stern}, {Ceverino}, {Dekel}, {Wang}, \& {Dange}}]{Strawn:2021ua}
{Strawn}, C., {Roca-F{\`a}brega}, S., {Mandelker}, N., {et~al.} 2021, \mnras,
  501, 4948.
\newblock \doarXiv{2008.11863}

\bibitem[{{Tan} \& {Oh}(2021)}]{Tan:2021tc}
{Tan}, B., \& {Oh}, S.~P. 2021, \mnras, 508, L37.
\newblock \doarXiv{2105.11496}

\bibitem[{{Terrazas} {et~al.}(2020){Terrazas}, {Bell}, {Pillepich}, {Nelson},
  {Somerville}, {Genel}, {Weinberger}, {Habouzit}, {Li}, {Hernquist}, \&
  {Vogelsberger}}]{Terrazas:2020wg}
{Terrazas}, B.~A., {Bell}, E.~F., {Pillepich}, A., {et~al.} 2020, \mnras, 493,
  1888.
\newblock \doarXiv{1906.02747}

\bibitem[{{Thompson} {et~al.}(2016){Thompson}, {Quataert}, {Zhang}, \&
  {Weinberg}}]{Thompson:2016vp}
{Thompson}, T.~A., {Quataert}, E., {Zhang}, D., \& {Weinberg}, D.~H. 2016,
  \mnras, 455, 1830.
\newblock \doarXiv{1507.04362}

\bibitem[{{Tomczak} {et~al.}(2014){Tomczak}, {Quadri}, {Tran}, {Labb{\'e}},
  {Straatman}, {Papovich}, {Glazebrook}, {Allen}, {Brammer}, {Kacprzak},
  {Kawinwanichakij}, {Kelson}, {McCarthy}, {Mehrtens}, {Monson}, {Persson},
  {Spitler}, {Tilvi}, \& {van Dokkum}}]{Tomczak:2014wi}
{Tomczak}, A.~R., {Quadri}, R.~F., {Tran}, K.-V.~H., {et~al.} 2014, \apj, 783,
  85.
\newblock \doarXiv{1309.5972}

\bibitem[{{Tripp} {et~al.}(2008){Tripp}, {Sembach}, {Bowen}, {Savage},
  {Jenkins}, {Lehner}, \& {Richter}}]{Tripp:2008tj}
{Tripp}, T.~M., {Sembach}, K.~R., {Bowen}, D.~V., {et~al.} 2008, \apjs, 177,
  39.
\newblock \doarXiv{0706.1214}

\bibitem[{{Tumlinson} {et~al.}(2013){Tumlinson}, {Thom}, {Werk}, {Prochaska},
  {Tripp}, {Katz}, {Dav{\'e}}, {Oppenheimer}, {Meiring}, {Ford}, {O'Meara},
  {Peeples}, {Sembach}, \& {Weinberg}}]{Tumlinson:2013vs}
{Tumlinson}, J., {Thom}, C., {Werk}, J.~K., {et~al.} 2013, \apj, 777, 59.
\newblock \doarXiv{1309.6317}

\bibitem[{{Voit}(2019)}]{Voit:2019tv}
{Voit}, G.~M. 2019, \apj, 880, 139.
\newblock \doarXiv{1811.04976}

\bibitem[{{Wakker} {et~al.}(2012){Wakker}, {Savage}, {Fox}, {Benjamin}, \&
  {Shapiro}}]{Wakker:2012vm}
{Wakker}, B.~P., {Savage}, B.~D., {Fox}, A.~J., {Benjamin}, R.~A., \&
  {Shapiro}, P.~R. 2012, \apj, 749, 157.
\newblock \doarXiv{1202.5973}

\bibitem[{{Werk} {et~al.}(2012){Werk}, {Prochaska}, {Thom}, {Tumlinson},
  {Tripp}, {O'Meara}, \& {Meiring}}]{Werk:2012ug}
{Werk}, J.~K., {Prochaska}, J.~X., {Thom}, C., {et~al.} 2012, \apjs, 198, 3.
\newblock \doarXiv{1108.3852}

\bibitem[{{Werk} {et~al.}(2013){Werk}, {Prochaska}, {Thom}, {Tumlinson},
  {Tripp}, {O'Meara}, \& {Peeples}}]{Werk:2013uj}
---. 2013, \apjs, 204, 17.
\newblock \doarXiv{1212.0558}

\bibitem[{{W}es {M}c{K}inney(2010)}]{McKinney:2010vw}
{W}es {M}c{K}inney. 2010, in {P}roceedings of the 9th {P}ython in {S}cience
  {C}onference, ed. {S}t\'efan van~der {W}alt \& {J}arrod {M}illman, 56 -- 61,
  \dodoi{10.25080/Majora-92bf1922-00a}

\bibitem[{{Whitaker} {et~al.}(2014){Whitaker}, {Franx}, {Leja}, {van Dokkum},
  {Henry}, {Skelton}, {Fumagalli}, {Momcheva}, {Brammer}, {Labb{\'e}},
  {Nelson}, \& {Rigby}}]{Whitaker:2014tq}
{Whitaker}, K.~E., {Franx}, M., {Leja}, J., {et~al.} 2014, \apj, 795, 104.
\newblock \doarXiv{1407.1843}

\bibitem[{{Wijers} {et~al.}(2020){Wijers}, {Schaye}, \&
  {Oppenheimer}}]{Wijers:2020vr}
{Wijers}, N.~A., {Schaye}, J., \& {Oppenheimer}, B.~D. 2020, \mnras, 498, 574.
\newblock \doarXiv{2004.05171}

\bibitem[{{Wilde} {et~al.}(2021){Wilde}, {Werk}, {Burchett}, {Prochaska},
  {Tchernyshyov}, {Tripp}, {Tejos}, {Lehner}, {Bordoloi}, {O'Meara}, \&
  {Tumlinson}}]{Wilde:2021vr}
{Wilde}, M.~C., {Werk}, J.~K., {Burchett}, J.~N., {et~al.} 2021, \apj, 912, 9.
\newblock \doarXiv{2008.08092}

\bibitem[{{Zahedy} {et~al.}(2019){Zahedy}, {Chen}, {Johnson}, {Pierce},
  {Rauch}, {Huang}, {Weiner}, \& {Gauthier}}]{Zahedy:2019vq}
{Zahedy}, F.~S., {Chen}, H.-W., {Johnson}, S.~D., {et~al.} 2019, \mnras, 484,
  2257.
\newblock \doarXiv{1809.05115}

\bibitem[{{Zinger} {et~al.}(2020){Zinger}, {Pillepich}, {Nelson}, {Weinberger},
  {Pakmor}, {Springel}, {Hernquist}, {Marinacci}, \&
  {Vogelsberger}}]{Zinger:2020vm}
{Zinger}, E., {Pillepich}, A., {Nelson}, D., {et~al.} 2020, \mnras, 499, 768.
\newblock \doarXiv{2004.06132}

\end{thebibliography}
\end{document}